\pdfoutput=1
\documentclass[a4paper]{article}
\usepackage{jheppub} 
\usepackage{graphicx}
\usepackage[english]{babel}
\usepackage{amssymb,amsfonts,amsmath}
\usepackage{verbatim}
\usepackage{physics}
\usepackage{cases}
\usepackage{mathtools}
\usepackage[dvipsnames]{xcolor}
\usepackage{tikz}
\usepackage{wrapfig}
\usepackage{subcaption}

\newcommand{\be}{\begin{equation}} \newcommand{\ee}{\end{equation}}
\newcommand{\bea}{\begin{eqnarray}} \newcommand{\eea}{\end{eqnarray}}
\DeclarePairedDelimiter\ceil{\lceil}{\rceil}

\newcommand{\nr}[1]{(\ref{#1})}  
\newcommand{\rmi}[1]{{\mbox{\scriptsize #1}}}  

\newcommand{\pt}{\partial}
\newcommand{\na}{\nabla}
\newcommand{\pminus}[1]{^{-#1}}

\newcommand{\reals}{\mathbb{R}}

\newcommand{\euc}{e}
\newcommand{\df}{\delta\Tilde{f}}

\newcommand{\nn}{\nonumber}
\newcommand{\G}{G_\text{N}}

\newcommand{\eps}{\epsilon}

\newcommand{\s}{\hspace{0.5pt}}
\newcommand{\p}{\partial}

\newcommand{\ccdot}{\,\cdot\,}

\makeatletter
\def\@fpheader{\relax}
\makeatother

\begin{document}

\title{Bulk metric reconstruction from entanglement data via minimal surface area variations}

\author[a,b]{Niko Jokela,}
\emailAdd{niko.jokela@helsinki.fi}
\affiliation[a,b]{Department of Physics and Helsinki Institute of Physics\\
P.O.~Box 64, FI-00014 University of Helsinki, Finland}

\author[c,d]{Tony Liimatainen,}
\emailAdd{tony.liimatainen@helsinki.fi}
\affiliation[c]{Department of 
Mathematics and Statistics, University of Helsinki, Finland}
\author[c]{Miika Sarkkinen,}
\emailAdd{miika.sarkkinen@helsinki.fi}
\affiliation[d]{Department of 
Mathematics and Statistics, University of Jyväskylä, Finland}

\author[e]{and Leo Tzou}
\emailAdd{leo.tzou@unimelb.edu.au}
\affiliation[e]{School of Mathematics and Statistics, The University of Melbourne, Australia}

\abstract{
We investigate the reconstruction of asymptotically anti-de Sitter (AdS) bulk geometries from boundary entanglement entropy data for ball-shaped entangling regions. By deriving an explicit inversion formula, we relate variations in entanglement entropy to deviations of the bulk metric about a fixed background. Applying this formula, we recover the AdS-Schwarzschild spacetime in the low-temperature regime to first order. We further extend our analysis to include deformations of the bulk geometry with nontrivial dependence on boundary directions, and propose an iterative reconstruction scheme aimed at recovering the full spacetime starting close to a conformal fixed point. We do this by building on recent advances in the mathematics of inverse problems by introducing the higher-order linearization method as a new tool in the context of holographic bulk reconstruction.
}

\preprint{HIP-2025-12/TH}

\maketitle

\setcounter{page}{2}

\newpage

\section{Introduction}\label{sec:intro}

The AdS/CFT correspondence is a profound duality that relates a gravitational theory in anti-de Sitter (AdS) spacetime to a conformal field theory (CFT) defined on its boundary. Within this framework, the degrees of freedom of the boundary CFT encode the dynamics of the bulk gravitational theory, providing a novel approach to studying strongly coupled quantum systems. This correspondence has had significant implications across various fields, including quantum chromodynamics, condensed matter physics, and quantum gravity~\cite{Brambilla:2014jmp,Ramallo:2013bua}.

One of the most striking features of AdS/CFT is its ability to translate complex problems in field theory into more tractable geometric problems in the bulk. A key example is the Ryu--Takayanagi(RT)~\cite{Ryu:2006bv} and Hubeny--Rangamani--Takayanagi~(HRT)~\cite{Hubeny:2007xt} proposals, which establish a connection between boundary entanglement entropy (EE) and extremal surfaces in the bulk. These proposals suggest that the entanglement structure of the boundary state plays a fundamental role in the emergence of bulk geometry. However, understanding how this boundary entanglement data reconstructs the full bulk spacetime remains an open problem in general.

Traditionally, holographic models of field theory phenomena begin with a well-defined supergravity action that exhibits clear symmetries. However, the reverse problem known as bulk reconstruction is equally intriguing and significantly more challenging. It involves reconstructing the dual holographic spacetime, along with any dynamic matter fields, from boundary field theory data. This process raises fundamental questions about how the boundary is represented within the bulk geometry. While various methods for bulk reconstruction have been proposed~\cite{Hamilton:2006az,Hammersley:2007ab,Bilson:2008ab,Bilson:2010ff,Kabat:2011rz,Balasubramanian:2013lsa,Spillane:2013mca,Headrick:2014eia,Czech:2015qta,Engelhardt:2016wgb,Hashimoto:2018ftp,Hernandez-Cuenca:2020ppu,Bao:2019bib,Hashimoto:2020mrx,Cao:2020uvb,Hashimoto:2021umd,Ahn:2024jkk}, most rely on precise knowledge of boundary quantities. In realistic scenarios, however, handling imprecise and discrete boundary data is crucial~\cite{Jokela:2020auu,Jokela:2023rba}.

Despite substantial progress, the question of how gravitational degrees of freedom emerge from a non-gravitational boundary theory remains one of the deepest mysteries in holography. Resolving this issue continues to drive research at the intersection of quantum field theory, gravity, and information theory.

In this work, we take a step toward addressing this challenge by seeking guidance from mathematics. Rather than assuming a specific bulk theory from the outset, we explore what mathematical structures can reveal about the possible ways in which a bulk spacetime could emerge from boundary data. While a complete classification of holographic spacetimes is beyond the scope of our study, our research premise is to aim to identify key mathematical principles that govern the reconstruction process. By doing so, we hope to gain insights into how boundary information constrains the possible geometric structures in the bulk and what this implies for the nature of holographic dualities. For instance, in gapped field theories, reconstructions using slab-shaped entangling regions fail because entanglement entropy saturates due to the finite correlation length~\cite{Nishioka:2006gr,Klebanov:2007ws,Jokela:2020wgs}. This is because of homogeneous boundary data; it would be interesting to attempt bulk reconstruction with more generic planar deformations~\cite{Faulkner:2015csl}. In contrast, ball-shaped entangling regions can probe the full bulk geometry, making them more suitable stepping stones for reconstruction. We see this work as a first step, laying the groundwork for future investigations that may eventually lead to a deeper understanding of the mathematical foundations of holography through entanglement.

We emphasize that key foundational questions remain unresolved. 
For example, AdS spacetime can clearly be foliated by maximally symmetric minimal surfaces, hemispheres of varying radii, hanging from the asymptotic boundary. Yet, it remains an open question whether a spacetime $\cal M$ with the same minimal surface areas as for AdS must necessarily be identified with AdS itself.
From a mathematical perspective, the difficulty of the bulk reconstruction problem is not surprising. The problem is related to inverse boundary value problems for partial differential equations (PDE), a field of mathematics with many long-standing open questions. The inverse problem most closely related to this work is the anisotropic Calder\'{o}n problem on Riemannian manifolds, which investigates whether a Riemannian manifold with boundary can be uniquely determined by the so called Dirichlet-to-Neumann map of the Laplace equation. The Dirichlet-to-Neumann map in general is an operator that sends a boundary value for a PDE to the normal derivative of the solution corresponding to the boundary value.  The connection of the anisotropic Calder\'{o}n problem to the  bulk reconstruction in AdS/CFT, is that area data of minimal surfaces determines the Dirichlet-to-Neumann map of the minimal surface equation.

Despite over four decades of active research in inverse problems, the anisotropic Calderón problem remains unresolved in dimensions three and higher, with only limited progress achieved in these cases \cite{uhlmann201330}. For example, even the fundamental question of whether a Riemannian manifold with the same boundary and Dirichlet-to-Neumann map as a Euclidean domain is Euclidean remains open.

Nevertheless, recent advances in inverse problems for \emph{nonlinear PDEs}, originating from \cite{kurylev2018inverse}, and subsequently for Laplace-type PDEs from \cite{feizmohammadi2020inverse,Carstea:2024npk}, have opened promising new research directions. For a recent survey, see \cite{lassas2025introductioninverseproblemsnonlinear}. These breakthroughs leverage nonlinearity as a powerful tool in inverse problems, with the \emph{higher-order linearization method} emerging as a key technique. This method involves taking multiple derivatives of the nonlinear equation and the associated Dirichlet-to-Neumann map with respect to small parameters in the data, enabling the recovery of various coefficients in inverse problems. Recently, this approach was employed in \cite{Carstea:2024npk} to reconstruct two-dimensional minimal surfaces from their areas on three-dimensional \emph{compact} Riemannian manifolds. This general method is discussed in detail in Section \ref{sec:compact_case}.

One of the primary objectives of this work is to establish a link between recent progress in the mathematics of inverse problems, particularly the results in \cite{Carstea:2024npk}, and the physics of the RT proposal, which relates areas of minimal surfaces to entanglement entropy. Specifically, we aim to adapt the higher-order linearization method to the study of the duality in asymptotically AdS spacetimes. With natural modifications, we anticipate that these methods will also find applications in the context of the HRT proposal for extremal surfaces.

\subsection{Problem statement and methodology}

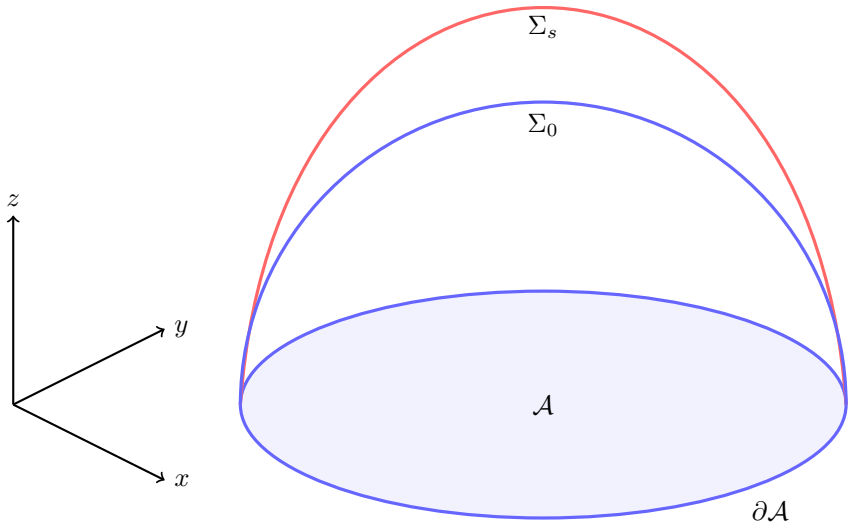
\begin{figure}
\begin{tikzpicture}
    \draw[thick,->] (0,0) -- (0,2.5) node[anchor=south]{$z$};
    \draw[thick,->] (0,0) -- (2,1) node[anchor=west]{$y$};
    \draw[thick,->] (0,0) -- (2,-1) node[anchor=west]{$x$};
    
    \filldraw[color=blue!60, fill=blue!5, very thick] (7,0) ellipse (4 and 1.5);
    \draw[color=red!60, very thick] (3,0) .. controls (3.5,7) and (10.5,7) .. (11,0);
    \draw[color=blue!60, very thick] (11,0) arc (0:180:4);

    \node at (7, 0) {${\cal A}$};
    \node at (7, 3.7) {$\Sigma_0$};
    \node at (7, 5) {$\Sigma_s$};
    \node at (10, -1.4) {$\partial \mathcal{A}$};
\end{tikzpicture}
\caption{Minimal surfaces anchored on the same boundary disk $\partial \mathcal{A}$ in pure AdS ($\Sigma_0$) and perturbed AdS ($\Sigma_s$). This deformation could arise from dialing some external parameter in the dual field theory or in an attempt to capture say contribution from additional degrees of freedom to the entanglement entropy such as flavors~\cite{Karch:2014ufa,Chalabi:2020tlw,Jokela:2024cxb}. Here $z$ is the holographic coordinate so that the boundary is at $z=0$.}\label{fig:RTembedding}
\end{figure}

Per holographic duality, the area $A$ of a minimal surface $\Sigma$ corresponds to the entanglement entropy of the associated boundary region ${\cal A}$~\cite{Ryu:2006ef},
\be\label{eq:RT}
 S_{EE} = \frac{A}{4\G} \ ,
\ee
where $\partial\Sigma=\partial{\cal A}$ and $\G$ is the Newton constant, which can be related with QFT quantities in string theory settings. See Fig.~\ref{fig:RTembedding} for an illustration. If one knows the metric of bulk spacetime, then it would be a straightforward exercise to compute the area $A$ of the codimension-two minimal surface $\Sigma$. The bulk reconstruction problem arises from knowing, in principle, the left-hand side of~\eqref{eq:RT}, the entanglement entropy, and seeking to determine the corresponding bulk geometry that would reproduce the area on the right-hand side. This constitutes an inverse problem, where one aims to infer geometric data from boundary entanglement information. 

However, it is important to recognize that the quantity in~\eqref{eq:RT} is formally divergent, and thus the entanglement entropy itself cannot directly serve as input data for reconstructing the bulk geometry. Instead, variations of the entanglement entropy relative to a reference configuration provide a well-defined and physically meaningful starting point. It is natural, then, to ask how does the entanglement entropy responds to changes in external parameters, or to deformations in the size or shape of the entangling region $\cal A$. These variations induce corresponding changes $\delta A$ in the areas of the associated bulk surfaces, thereby encoding information about the underlying geometry.

We investigate the problem of recovering the metric of a spacelike hypersurface from the area data of minimal surfaces anchored at infinity of asymptotically AdS spacetime. We restrict our analysis to the recovery of the function $f$ in the metric:
\begin{equation}\label{eq:metric_intro}
g(z,x) = \frac{1}{z^2}\left( f(z,x) \, \dd z^2 + e \right),
\end{equation}
where $e$ is the Euclidean metric and $x=(x_1,x_2,\ldots, x_{n-1}) \in \mathbb{R}^{n-1}$. Metrics of this form include, for example, time slices of AdS black holes (see Section \ref{sec:linear}). While the methods of \cite{Carstea:2024npk} demonstrate that $f$ can be recovered in the compact setting (and for $n = 3$), they rely on the existence of complex geometric optics (CGO) solutions, or Faddeev's Green's functions. Unfortunately it is unclear what the corresponding solutions are in the noncompact setting. For this reason, we focus on recovering perturbations $\delta f$ about pure AdS, {\emph{i.e.}}, $f = 1+\delta f$, utilizing known solutions to the linearized minimal surface equation in pure AdS \cite{Hubeny:2012ry}.

A perturbation $\delta f$ of the metric \eqref{eq:metric_intro} leads to a perturbation $\delta A$. Here, \( \delta A \) is considered a function of how minimal surfaces are anchored to infinity \( z = 0 \). For example, if minimal surfaces are anchored to a boundary of a ball (a sphere) of radius $R$, we denote the bulk area perturbation by $ \delta A(R)$. This $\delta A(R)$ is supplemented to us by the entanglement entropy data about a ball-shaped subregion either by changes in external parameters (temperature) or responses to area deformations such as scaling.
In the latter case, this means to have access to derivatives $\delta S_{EE}'(R),\delta S''_{EE}(R),\ldots$, which translates to $\delta A'(R), \delta A''(R),\ldots$. Void of any other length or energy scale, one is bound to making a deformation of the entangling region, scaling of the radius being the simplest one. 

One main focus in this work is that we indeed allow general shape deformations. We discuss this in Section~\ref{sec:linear} in the context of bulk reconstruction. See also other interesting work for shape deformations of balls within holography \cite{Faulkner:2015csl,Mezei:2014zla}.
It is worth emphasizing that in this work we focus on entangling regions which are not strips nor slabs that have thus far been the most popular ones within bulk reconstruction. We consider entangling regions which are close to balls. 

At first glance, it might seem peculiar that our starting point involves hemisphere embeddings, the minimal surfaces in pure AdS. However, this choice is natural when considering small entangling regions. In asymptotically AdS spacetimes, the corresponding minimal surfaces for small regions remain close to those in pure AdS. Since we are perturbing around AdS, there is no intrinsic scale against which ``small" can be measured within the deformed geometry itself. Therefore, it is both convenient and justified to begin with the pure AdS background and parametrize deviations via 
$f(z)=1+\delta f(z)$. In this way, we reconstruct the geometry near the boundary and propose a scheme to recover the full bulk geometry beyond leading order.

We do not impose any assumptions on the equations satisfied by the bulk metric (other than requiring it to take the form \eqref{eq:metric_intro}). However, we would like to highlight a connection to works such as \cite{Lashkari:2013koa,Faulkner:2013ica,Swingle:2014uza}, which, as we do, study the bulk reconstruction problem on the perturbative level. There $\delta S_{EE}$  is assumed not only to encode perturbations in the areas of minimal surfaces under the RT proposal but also to satisfy
\begin{equation}\label{eq:hyperbolic_energy}
\delta S_{EE} = \delta E \ ,
\end{equation}
where $\delta E$ is the ``hyperbolic" energy of the perturbed state. It has been shown \cite{Blanco:2013joa} that if the bulk metric obeys the linearized vacuum Einstein equations, then \eqref{eq:hyperbolic_energy} holds. Thus, this condition can be viewed as a holographic dual of the Einstein equations (at the perturbative level) in the vacuum. Moreover, \cite{Lashkari:2013koa} established that this condition is both necessary and sufficient under broad assumptions. While their work does not reconstruct the bulk metric under the additional constraint \eqref{eq:hyperbolic_energy}, they demonstrate that the bulk metric must satisfy the linearized Einstein equations. 

In the mathematics literature, the related Calder\'on problem (see the earlier discussion) has been studied for Einstein metrics in \cite{guillarmou2007inverse}. Although their results are not directly applicable to the RT proposal (particularly because the linearized minimal surface equation is a Schr\"odinger  equation rather than the Laplace equation), their analysis suggests that bulk reconstruction from EEs should be feasible in general settings for Einstein metrics (of Euclidean signature). To our knowledge, the RT bulk reconstruction problem under \eqref{eq:hyperbolic_energy} has not been studied in the mathematical literature, making it a compelling open question.

\subsection{Summary of results}

\subsubsection{Recovery of $\delta f(z)$}

We first consider the case where \( \delta f = \delta f(z) \), {\emph{i.e.}}, \( \delta f \) depends only on \( z \) and $n\geq 3$. We show that \( \delta f \) can be recovered from the area data of minimal surfaces anchored at infinity as spheres \( S^{n-2} \) of varying radii $R$. For \( n = 3 \), we derive the explicit inversion formula:
\begin{equation}\label{eq:inversion_intro}
\delta f(z) = \frac{12\G}{\pi}\left( -\delta S_{EE}(z) + z\, \delta S_{EE}'(z) + z^2 \delta S_{EE}''(z) \right) \ 
\end{equation}
holding for $z>0$. 
This formula indicates that recovering $\delta f(z)$ at ``height" $z = R$ requires knowledge of $\delta S_{EE}(R)$ and the first and second derivatives, \( \delta S_{EE}'(R) \) and \( \delta S_{EE}''(R) \).   We validate this method for the AdS black hole metric:
\begin{equation}\label{eq:BH_metric_intro}
f(z) = \frac{1}{1 - (z/z_H)^3} = 1 + \left(\frac{z}{z_H}\right)^3 + \left(\frac{z}{z_H}\right)^6 + \ldots \ ,
\end{equation}
recovering the leading-order perturbation \( \delta f = (z/z_H)^3 \) from \( \delta S_{EE}(R) \). We derive similar formulas $\delta f(z)$ as \eqref{eq:inversion_intro} for \( n > 3 \), through higher-order linearizations to order \( n+1 \) or \( n \), depending on whether \( n \) is odd or even, of \( \delta S_{EE}(R) \) are required for them.

\subsubsection{Recovery of $\delta f(z, x, y)$}

Next, we study the case where \( \delta f = \delta f(z, x, y) \) for \( n = 2 \). We assume \( \delta f \) admits a finite-order spherical harmonic expansion:
\[
\delta f(\rho, \theta, \phi) \approx \sum_{l=0}^k \sum_{m=-l}^l a_{lm}(\rho) Y_l^m(\theta, \phi)\ ,
\]
and derive a general second-order linearization integral identity for \( \delta A \), schematically written as:
\begin{equation}\label{integral-id_intro}
\frac{d}{d\epsilon_2}\frac{d}{d\epsilon_1}\Big\rvert_{\epsilon_1=\epsilon_2=0}\delta A_{\epsilon_1,\epsilon_2} = \int \left( H_1 v_1 v_2 + H_2(\nabla v_1, \nabla v_2) + H_3 \cdot (v_1 \nabla v_2 + v_2 \nabla v_1) + H_4 w + H_5 \cdot \nabla w \right)\ .
\end{equation}
This identity corresponds to varying how the minimal surface is anchored to infinity with respect to two parameters \( \epsilon_1 \) and \( \epsilon_2 \). This quantity can be considered as the linearized version of entanglement density $\partial_{\epsilon_2}\partial_{\epsilon_1}|_{\epsilon_1=\epsilon_2=0} A_{\epsilon_1,\epsilon_2}$ of \cite{Nozaki:2013wia, Bhattacharya:2014vja}. Here, \( H_1, \ldots, H_5 \) are functions/tensors that depend on \( \delta f(z, x, y) \), while \( v_1, v_2 \), and \( w \) are solutions to the first and second linearized minimal surface equation in pure AdS. We expand the coefficients \( H_1, \ldots, H_5 \) in terms of the spherical harmonic coefficients $a_{lm}$ of \( \delta f \). Then, by plugging in a special set of solutions $v_1$, $v_2$, and $w$ into \eqref{integral-id_intro}, we extract information about the coefficients. This is a typical argument in inverse problems for PDEs.

The special solutions we use for \( v_1 \) and \( v_2 \) are provided by Hubeny in \cite{Hubeny:2012ry}. These solutions take the form \( \Phi(\phi)\Theta(\theta) \), where:
\[
\Phi(\phi) = \text{linear combination of } \sin (m\phi) \text{ and } \cos (m\phi),
\]
\[
\Theta(\theta) = (1 + m \cos \theta)\left(\frac{1 - \cos \theta}{1 + \cos \theta}\right)^{m/2},
\]
with \( m \in \mathbb{Z} \) as a free parameter. By selecting specific values of \( m \) for \( v_1 \) and \( v_2 \), we solve for the spherical harmonic coefficients up to order \( l = 4 \). Recovering coefficients for a general spherical harmonic expansion by this method remains an open question, but we note that using Hubeny's solutions in \eqref{integral-id_intro} provides infinitely many conditions for the coefficients of the expansion.

\subsection{Recovering $f$ from $A$ using inversion of $\delta A$}
Finally, we address the reconstruction of not just the perturbation $\delta f(z)$, but the \emph{entire} metric function $f(z)$, under the assumption that $f$ remains close to unity ($|f(z) - 1| \ll 1$). This reconstruction is enabled by the explicit inversion formula \eqref{eq:inversion_intro} in dimension $3$ and its higher-dimensional generalizations \eqref{eq: odd inversion} and \eqref{eq: even inversion}.

The statement of the problem is the following: given the minimal surface area function $r \mapsto A_f(r)$, associated with an unknown metric function $f \approx 1$, we aim to reconstruct $f(z)$ entirely from the boundary data $A_f$. The key insight is the first-order approximation
\begin{equation}\label{eq:approx_with_square_error}
A_f - A_{f=1} = \delta A + {\cal{O}}(|f - 1|^2) \ ,
\end{equation}
where $\delta A$ is the linearized area variation induced by the perturbation $\delta f = 1 - f$. Since $A_f - A_{f=1}$ is known from boundary measurements, the inversion formula for $\delta f$ yields a first-order approximation of $1 - f(z)$ in terms of this observable difference.

While \eqref{eq:approx_with_square_error} captures the leading-order behavior, higher-order corrections ${\cal{O}}(|f - 1|^2)$ must be incorporated for an accurate reconstruction. To achieve this, we propose an iterative scheme inspired by the implicit function theorem:
\begin{enumerate}
    \item \textbf{Initialization}: Begin with the trivial AdS metric $f_0(z) \equiv 1$.
    \item \textbf{Iterative Update}: At each step $n$, refine the estimate via
    \begin{equation}
    f_{n+1}(z) = f_n(z) - (D A|_{f=1})^{-1}(A_{f_n} - A_f) \ ,
    \end{equation}
    where $(D A|_{f=1})^{-1}$ is the inverse operator provided by our explicit linearized inversion formula \eqref{eq:inversion_intro}, $A_{f_n}$ is the area data for minimal surfaces computed for $f_n$, and $A_f$ is the given area data.
    \item \textbf{Convergence}: The iteration proceeds until $f_n$ converges, reconstructing the full nonlinear metric $f(z)$.
\end{enumerate}

This method is particularly suited for low-temperature black hole metrics \eqref{eq:BH_metric_intro} (where $z_H \gg 1$), as they remain perturbatively close to AdS. We discuss convergence criteria, numerical implementation, and broader implications in Section~\ref{sec:nonlinear}.

\section{Bulk reconstruction}\label{sect: setup}

In this work, we investigate the problem of \emph{recovering bulk geometric deformations} from corresponding \emph{variations in boundary entanglement entropy}. Specifically, we address the bulk reconstruction problem at the linearized (perturbative) level, working to first order around a fixed background geometry. In this regime, changes in entanglement entropy translate into variations in the areas of extremal surfaces, which in turn encode information about the underlying bulk geometry.

To make this connection precise, we consider small perturbations of the bulk geometry around a fixed background metric $g_0$, typically taken to be pure AdS unless otherwise specified. That is, we write the full metric as  
\be
    g = g_0 + \delta g \ ,
\ee
where $\delta g$ encodes the leading-order deviation from the background. According to the RT proposal, the entanglement entropy of a boundary region $\mathcal{A}$ in the CFT is dual to the area of a codimension-two minimal surface $\Sigma$ in the bulk satisfying $\partial\Sigma = \partial\cal{A}$.
In the perturbative setting, this translates into a variation of the minimal surface area,
\be
    A = A_0 + \delta A \ ,
\ee
where $A_0$ is computed in the background geometry and $\delta A$ captures the leading-order change due to $\delta g$. A more general, nonlinear, setting is discussed in Section~\ref{sec:nonlinear}.

Given the knowledge of how entanglement entropy changes under small deformations of the entangling region or under variations of external parameters such as temperature, the RT formula allows us to relate these variations to the area corrections via $\delta A = 4G \, \delta S_{\text{EE}}$. The central inverse problem we address is to reconstruct the metric perturbation $\delta g$ from this area data.

In Section~\ref{sec:linear}, we focus on spherical boundary regions, where $\partial\Sigma$ are spheres of radii $r>0$ at the boundary of AdS. It is worth emphasizing that we keep the boundary regions intact, so the entanglement data that induces area perturbations stem from varying external parameters. Using these, we reconstruct perturbations $\delta f(z)$ of the function $f = 1$ in the metric
\begin{equation}
    g = \frac{1}{z^2}\left(f(z)\,\dd z^2 + e\right),
\end{equation}
where $f(z)$ depends only on the holographic coordinate $z$.

\begin{figure}
\subfloat[$k=0$]{\includegraphics[width = 0.33\linewidth]{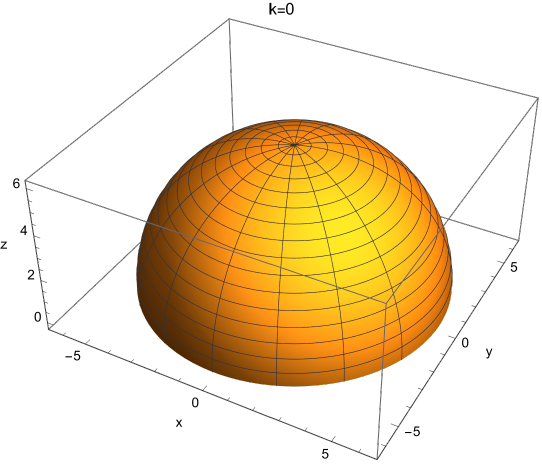}} 
\subfloat[$k=1$]{\includegraphics[width = 0.33\linewidth]{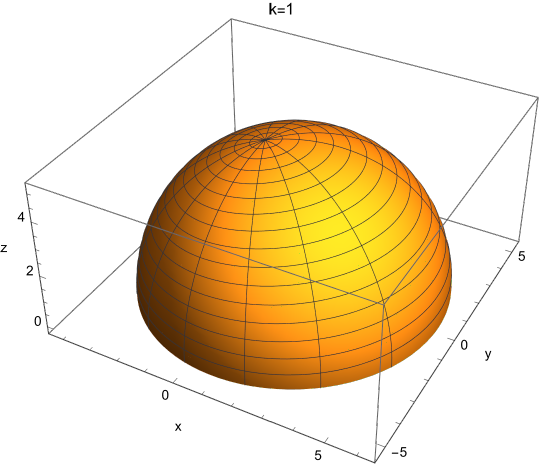}}
\subfloat[$k=2$]{\includegraphics[width = 0.33\linewidth]{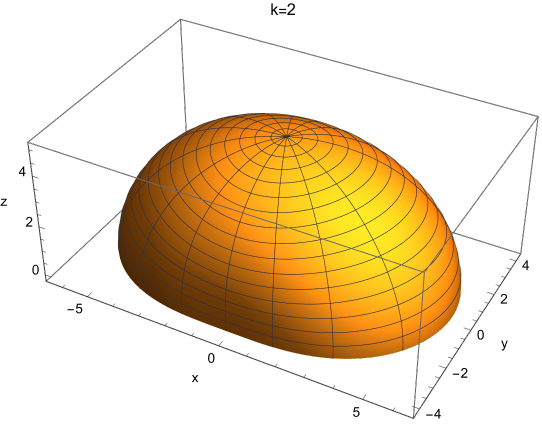}}\\
\subfloat[$k=3$]{\includegraphics[width = 0.33\linewidth]{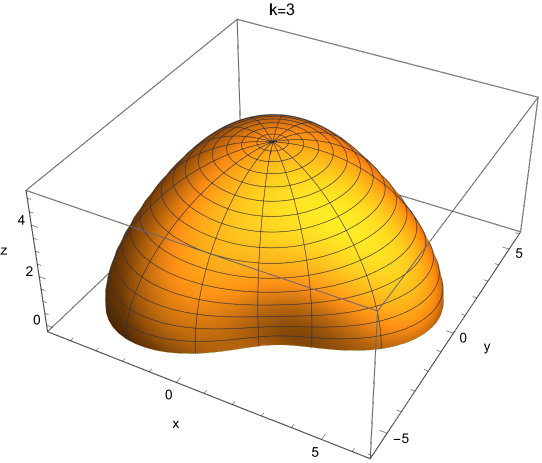}}
\subfloat[$k=4$]{\includegraphics[width = 0.33\linewidth]{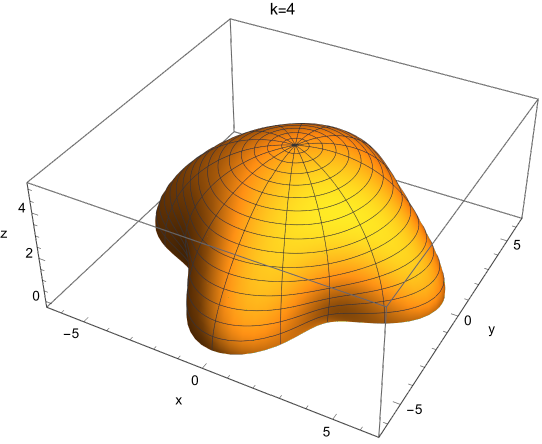}}
\subfloat[$k=5$]{\includegraphics[width = 0.33\linewidth]{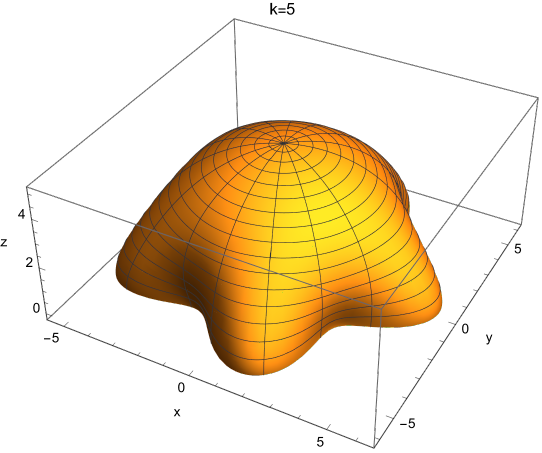}}
\caption{Linearized solutions $u_k$ to the minimal surface equation around the exact hemisphere solution in pure AdS.}
\label{fig:example}
\end{figure}

In Section~\ref{sec:shapes}, we generalize this approach by deforming spherical boundaries using trigonometric functions, using solutions by Hubeny \cite{Hubeny:2012ry} (see Fig.~\ref{fig:example}). This allows us to give an example of a reconstruction of perturbations $\delta f(z,x,y)$ of the function $f = 1$ in the metric
\begin{equation}
    g = \frac{1}{z^2}\left(f(z,x,y)\,\dd z^2 + e\right),
\end{equation}
now depending on all boundary coordinates $(x, y)$ as well as $z$.

Throughout the paper, we employ the spherical coordinate system adapted to the symmetries of the problem, which make explicit how the radius $r$ of the boundary region determines the minimal surface embedding.

\subsection{Review of bulk recovery in the compact case and the higher-order linearization method}\label{sec:compact_case}
As said, we address the bulk reconstruction problem at the perturbative level. Our approach is however motivated by recent developments in the nonlinear problem in the compact case, where the spacelike
hypersurface is a compact Riemannian manifold (as opposed to an asymptotically AdS spacetime).

When the spacelike hypersurface is a compact Riemannian manifold, the recovery of the metric from the areas of minimal surfaces anchored at its boundary has been recently explored in \cite{alexakis2020determining,Carstea:2024npk} (see also \cite{Bao:2019bib}). 
The works \cite{alexakis2020determining,Carstea:2024npk} consider the full nonlinear problem of reconstructing the bulk metric from minimal surface area data, and mostly focus on the case where the Riemannian manifold is three dimensional and the minimal surfaces are two-dimensional.  
The first of these works \cite{alexakis2020determining} recovers the three-dimensional bulk metric under additional geometric and topological assumptions.  The second \cite{Carstea:2024npk} recovers individual two-dimensional minimal surfaces in a general setting but does not address how these surfaces can be ``glued" together to reconstruct the full three-dimensional bulk metric.

It remains an open question whether the methods developed in \cite{Carstea:2024npk, alexakis2020determining} can be extended to non-compact Riemannian manifolds with asymptotic infinity. 
Our linearized approach avoids the potential complications of these methods in the non-compact setting while still providing physically relevant insights into the bulk reconstruction problem. The restriction to first-order perturbations also allows us to develop reconstruction methods that are also computationally tractable.
To provide context and motivation for our approach, we next outline how metric recovery is achieved in the compact case and under what conditions. This discussion naturally leads into the methodology employed in the present work.

\begin{wrapfigure}{r}{7cm}
\begin{tikzpicture}
\fill[gray!20] plot[smooth] coordinates { (2,0.5) (3.,1.1) (3,2) } -- plot[smooth cycle, tension=0.7] coordinates { (0,0) (2,0.5) (3,2) (2.5,3.5) (1,4) (-1,3.5) (-2.5,2) (-2,0.5) };
\fill[gray!20] plot[smooth] coordinates { (-2,0.5) (-2.9,1.35) (-2.5,2) } -- plot[smooth cycle, tension=0.7] coordinates { (0,0) (2,0.5) (3,2) (2.5,3.5) (1,4) (-1,3.5) (-2.5,2) (-2,0.5) };
\fill[gray!20] plot[smooth] coordinates { (-0.5,0.02) (0.5,-0.3) (1,0.08) } -- plot[smooth cycle, tension=0.7] coordinates { (0,0) (2,0.5) (3,2) (2.5,3.5) (1,4) (-1,3.5) (-2.5,2) (-2,0.5) };

 \draw[dashed,thick] plot[smooth] coordinates { (2,0.5) (3.,1.1) (3,2) };
 \draw[dashed,thick] plot[smooth] coordinates { (-2,0.5) (-2.9,1.35) (-2.5,2) };
 \draw[dashed,thick] plot[smooth] coordinates { (-0.5,0.02) (0.5,-0.3) (1,0.08) };

\draw[thick] plot[smooth cycle, tension=0.7] coordinates { (0,0) (2,0.5) (3,2) (2.5,3.5) (1,4) (-1,3.5) (-2.5,2) (-2,0.5) };

\node at (3.3, 1.4) {$F_1$};
\node at (-3.05, 1.8) {$F_3$};
\node at (1.1, -0.2) {$F_2$};

\node at (-0.2, 4.1) {$\partial {\cal A}$};

\node at (0, 2.1) {${\cal A}$};

\end{tikzpicture}
\caption{Illustration of the boundary deformation $\delta(\partial{\cal A})$ of the entangling region $\cal A$ enclosed by the black curve $\partial{\cal A} = \partial\Sigma$. We will limit on deforming $\partial{\cal A}=S^1$, however.}\label{fig:bumbs}
\end{wrapfigure}
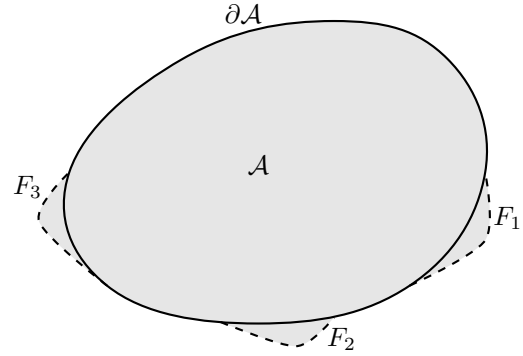
The recovery of minimal surfaces from the area data of minimal surfaces in \cite{Carstea:2024npk} relies on the
\emph{higher-order linearization method}, a technique recently developed in the study of inverse problems for nonlinear partial differential equations \cite{kurylev2018inverse,feizmohammadi2020inverse,lassas2021inverse}. 
To explain the method in more detail,  suppose that $X$ is a manifold with boundary $\partial X$ equipped with a Riemannian metric $g$. Consider also an embedded minimal surface $\Sigma$ whose boundary $\p \Sigma$ lies entirely in $\partial X$. 

Next consider nearby minimal surfaces, which can be considered as solutions $u=u^\mu$ to the minimal surface equation (see e.g. \cite[Appendix B.2]{Hubeny:2007xt})

\begin{equation}\label{eq:B5}
g^{\alpha\beta}\,\left(
\partial_{\alpha} \partial_{\beta} u^{\mu} + 
\Gamma^{\mu}_{\nu \lambda} \, \partial_{\alpha} u^{\nu} \, \partial_{\beta} u^{\lambda} - 
\Gamma^{\gamma}_{\alpha \beta} \, \partial_{\gamma} u^{\mu}\right)=0 \ ,
\end{equation}
where $\alpha,\beta$ correspond to coordinates on the surface and the other Greek indices go from $1,\ldots, \text{dim}(X)$.
We let the nearby solutions anchor to $\p X$ in a manner, which is described by three deformations $F_1, F_2$, and $F_3$ of $\p \Sigma$ and their magnitudes $\eps_1, \eps_2$, and $\eps_3$. This gives us a three parameter family of minimal surfaces, which we write as
\[
u_{\eps_1F_1+\eps_2F_2+\eps_3F_3} \ .
\]
Each $u_{\eps_1F_1+\eps_2F_2+\eps_3F_3}$ is the unique solution to the minimal surface equation determined by the deformation $\eps_1F_1+\eps_2F_2+\eps_3F_3$ of $\p\Sigma$, illustrated in Fig.~\ref{fig:bumbs}, and $\eps_1=\eps_2=\eps_3=0$ corresponds to the original minimal surface $\Sigma$ with boundary $\p\Sigma$.
In \cite{Carstea:2024npk}, the minimal surfaces are two-dimensional, so the functions $F_j$ are defined on  one-dimensional manifolds. In the current work, the functions $F_j$ can for example be defined on the circle $S^1$, which corresponds to the intersection of a hemisphere with the asymptotic infinity. See  Section \ref{sec:scaling}. (To avoid possible confusion, we  mention that \cite{Carstea:2024npk} uses Fermi-coordinates to write the minimal surface equation \eqref{eq:B5}.)

Let us use the notation $\eps=0$ to denote $\eps_1 = \eps_2= \eps_3=0$.  By taking the derivative $\p_{\eps_j}|_{\eps=0}$ of the minimal surface $u_{\eps_1F_1+ \eps_2F_2 + \eps_3F_3}$, 
we have that the function
\be
 v^j:=\frac{\p}{\p \eps_j}\Big|_{\eps=0}\s\s u_{\eps_1F_1+ \eps_2F_2 + \eps_3F_3}
\ee
solves the linearized minimal surface equation
\begin{equation}\label{eq:linearized_equation}
 \Delta_gv+qv=0 \ 
\end{equation}
(assuming in addition that the variation in $\eps_j$ is normal, {\emph{i.e.}}, $g(v_j,T\Sigma)=0$).
Above, $g=g_{\alpha\beta}$ is the induced metric on the minimal surface (which we aim to recover), and $q$ is a certain geometry dependent function on the minimal surface. If the area data of two-dimensional minimal surfaces is known, it is possible to recover 
$g$ up to a conformal transformation by studying solutions to the linearized equation \eqref{eq:linearized_equation} as was shown in~\cite{Carstea:2024npk}. In other words, the first linearization allows us to recover the minimal surface up to a conformal transformation. (In fact, this recovery corresponds to a solution~\cite{Carstea:2024npk} to a  Calder\'on inverse problem on Riemann surfaces.) Due to the coordinate invariance of the problem, the recovery of $g$ is, at best, possible only up to an isometry.

To recover the remaining unknown conformal factor, the argument  in the compact case proceeds to higher-order linearizations. By applying the area data to the second linearizations
\be
\frac{\p^2}{\p \eps_j\p \eps_k}\big|_{\eps=0}\s\s u_{\eps_1F_1+ \eps_2F_2 + \eps_3F_3} \ ,
\ee
one can recover quantities such as the extrinsic curvature, {\emph{i.e.}}, the second fundamental form. 
Roughly speaking, the recovery of the quantities here relies on demonstrating that products of three complex geometric optics (CGO) solutions, along with their gradients, to the linearized equation \eqref{eq:linearized_equation} form a complete set. Once the coefficients of the second linearization of the minimal surface equation are determined, one proceed to the third linearization. The conformal factor can then be recovered from the third linearization and the area data. See~\cite{Carstea:2024npk} for these steps. 

To construct the full three-dimensional bulk metric, the recovered two-dimensional minimal surfaces must still be glued together. 
For this, one at least needs to assume that all of the bulk can be reached by some minimal surfaces, which are anchored on the boundary entangling regions.\footnote{We note that holographic geometries have so called entanglement shadows~\cite{Balasubramanian:2014sra,Freivogel:2014lja}, which are regions which cannot be probed by any minimal or extremal surfaces. The study of such geometries may require additional input beyond spatial entanglement data~\cite{Balasubramanian:2017hgy}.}
In \cite{alexakis2020determining}, this gluing was achieved under specific curvature and foliation assumptions. 

In summary, the recovery process of the bulk metric in the compact case proceeds as follows:
\begin{enumerate}
    \item First linearization: Recover the minimal surface up to a conformal transformation.
    \item Second linearization: Solve the second linearized problem to prepare for the third linearization.
    \item Third linearization: Recover the conformal factor, completing the recovery of the minimal surface up to an isometry. 
    \item Gluing: Combine the recovered minimal surfaces to construct the full 3-dimensional bulk metric.
\end{enumerate}
We anticipate that these steps can ultimately be extended to the non-compact case as well. It is worth noting that the solutions to each of these steps are not very explicit.

To the best of our knowledge, it remains unknown how any of the four steps outlined above can be accomplished when the spacelike hypersurface under study is non-compact and possesses an asymptotic infinity. In this work, we demonstrate that these steps can be applied at the perturbative level in certain physically motivated special cases. Due to the special form of the metric we study (see \eqref{ads-metric} below) the isometry in the recovery is in fact just the identity.
Interestingly, at the perturbative level, we also find that the steps outlined above can have surprisingly explicit solutions.

\subsection{References to mathematical literature}
Recent mathematical developments in bulk reconstruction include the work ~\cite{Marx-Kuo:2024xmu}, which reconstructs the bulk metric from renormalized areas of minimal surfaces in asymptotically hyperbolic manifolds by ``Taylor'' expanding the metric at asymptotic boundary. In compact settings, inverse problems for minimal surfaces have been investigated in various geometric contexts~\cite{CARSTEA_toy_model,Nurminen_2023,nurminen2}, including Euclidean space $\mathbb{R}^n$. The Calder\'on problem on hyperbolic manifolds was considered by~\cite{isozaki2004inverse}. In the AdS/CFT context,~\cite{Alexakis:2010zz} established connections between renormalized areas of minimal surfaces and the Dirichlet-to-Neumann map for the (linearized) minimal surface equation.

\section{Linearized problem retaining Poincar\'e invariance}\label{sec:linear}

\subsection{Derivation of the minimal surface equation}

In this subsection we derive the minimal surface equation on a time slice of an asymptotically AdS spacetime.
We consider an asymptotically AdS$_{n+1}$ metric of the form
\begin{equation}\label{ads-metric}
    g_{\text{AdS}} = \frac{l^2}{z^2}\left( - \frac{1}{f(z)} \dd t^2 + f(z) \dd z^2 + \euc\right) ,
\end{equation}
where $f$ is an arbitrary function satisfying $f(0) = 1$, $l$ is the radius of curvature, and $\euc$ is the flat Euclidean metric in $\mathbb{R}^{n-1}$. Notice that we are assuming homogeneity and isotropy to start with.
The induced metric on spacelike hypersurfaces of constant $t$ is then given by
\begin{equation}\label{t-slice}
    g = \frac{l^2}{z^2}\left( f(z) \dd z^2 + \euc\right) .
\end{equation}
This is an asymptotically hyperbolic Riemannian metric on a manifold $M$ of dimension $n$. Conjoining the conformal infinity of AdS to the bulk manifold, we can think of $M$ as a manifold with a boundary separated by infinite proper distance from the points in the bulk. The boundary corresponds to the value of the holographic coordinate $z=0$.

Now consider an embedded hypersurface $\Sigma \subset M$ such that $\pt \Sigma \subset \pt M$. Let $x=(x^i), i=1,\ldots,n-1$ be the boundary coordinates of $M$ and parametrize the hypersurface in terms of $x$; the embedding of $\Sigma$ then is $F(x) = (x,z(x))$. The chosen embedding works at least locally, points where $\na z =(\pt_{x^1} z,...,\pt_{x^{n-1}}z)$ becomes singular have to be dealt with separately. The differential map of $F$ is 
\begin{equation}
    \dd F(x) = \begin{bmatrix}
        I_{(n-1)\times (n-1)} \\
        \na z (x) \\
    \end{bmatrix}
\end{equation}
so that the coordinate basis vectors are mapped as
\begin{equation}
    \dd F(\pt_{x^i}) = \pt_{x^i} + \frac{\pt z}{\pt x^i} \pt_z.
\end{equation}
The induced metric on $\Sigma$ then reads
\begin{equation}
    g_\Sigma = F^* g = g(\dd F(\pt_{x^i}),\dd F(\pt_{x^j})) \dd x^i \dd x^j = \frac{l^2}{z^2} \left(
    f(z) \frac{\pt z}{\pt x^i}\frac{\pt z}{\pt x^j}\dd x^i \dd x^j + \euc 
    \right) \ .
\end{equation}
The area of $\Sigma$ is obtained by integrating the induced volume form over the submanifold
\begin{equation}
    A_\Sigma = \int_{x(\Sigma)} \sqrt{\det (g_\Sigma)} \dd x^1 \wedge \cdots \wedge \dd x^{n-1}\ ,
\end{equation}
where the induced metric determinant is given by the formula
\begin{equation}
    \det (g_\Sigma) = \frac{l^{2(n-1)}}{z^{2(n-1)}}\det(\euc) \det\left( I_{(n-1)\times(n-1)} + f(z) \euc\pminus{1}(\cdot, \na z) \otimes \na z \right) \ .
\end{equation}
With Sylvester's determinant identity
\begin{equation}
    \det\left( I + \Vec{v} \otimes \Vec{w}^T \right) = 1 + \Vec{w}^T \cdot \Vec{v} \ ,
\end{equation}
we simply get
\begin{equation}
    \det (g_\Sigma) = \frac{l^{2(n-1)}}{z^{2(n-1)}} (1 + f(z) \abs{\na z}^2_{\euc}) \det(\euc) \ ,
\end{equation}
where $\abs{\cdot}_e$ is the norm with respect to the Euclidean metric.
The hypersurface area then becomes
\begin{equation}\label{eq:area-function}
    A_\Sigma = l^{n-1} \int_{x(\Sigma)} \frac{\sqrt{1 + f(z(x^i))\abs{\na z(x^i)}^2_{\euc}}}{z(x^i)^{n-1}}\sqrt{\det(\euc)}\dd x^1 \wedge \cdots \wedge \dd x^{n-1}. 
\end{equation}
With AdS asymptotics this is a divergent quantity, which from a physical point of view must be understood in a renormalized sense. The Euler-Lagrange equation for the area functional reads
\begin{align}
    0 &= \frac{\pt L}{\pt z} - \frac{\pt}{\pt x^i} \left( \frac{\pt L}{\pt(\pt_i z)} \right) \nn \\
   &= \sqrt{\det(\euc)}\left(\frac{f'(z)\abs{\na z}^2_{\euc}}{2 z^{n-1} \sqrt{1 + f(z)\abs{\na z}^2_{\euc}}} - (n-1) \frac{\sqrt{1 + f(z)\abs{\na z}^2_{\euc}}}{z^n} \right) \nn \\
   & \quad - \frac{\pt}{\pt x^i} \left(\sqrt{\det(\euc)} \frac{f(z) e^{ij}\pt_j z}{z^{n-1} \sqrt{1 + f(z)\abs{\na z}^2_{\euc}}} \right)\ .
\end{align}
Dividing by $\sqrt{\det(\euc)}$, we can write the divergence term in covariant form
\begin{equation}\label{eq:nonlinear-minsurf}
    \frac{z f'(z)\abs{\na z}^2_{\euc}-2(n-1)(1+f(z)\abs{\na z}^2_{\euc})}{2 z^{n} \sqrt{1 + f(z)\abs{\na z}^2_{\euc}}} - \text{div}_{\euc} \left( \frac{f(z) e\pminus{1}(\na z,\cdot)}{z^{n-1} \sqrt{1 + f(z)\abs{\na z}^2_{\euc}}} \right) = 0 \ .
\end{equation}
This is the full nonlinear minimal surface equation, which is in general very difficult to solve. Some useful closed-form solutions are known in pure AdS where $f=1$, \emph{e.g.}, an $(n-1)$-dimensional hemisphere with the equator glued to an $(n-2)$-dimensional sphere on the boundary \cite{Hubeny:2012ry}.

Ideally, we would like to recover the unknown function $f(z)$ assuming that we know the associated minimal surface areas based on entanglement entropy data on the boundary. An explicit inversion is feasible at least when the boundary entangling region is formed by an infinite strip/slab \cite{Bilson:2010ff}. However, as discussed in Introduction, there are severe limitations to bulk reconstruction using strip-like configurations. To bypass these complications, we shall consider the simplest possible smooth and bounded entangling regions, \emph{i.e.}, balls. In this case an explicit inversion of $f(z)$ via methods of \cite{Bilson:2010ff,Jokela:2020auu,Hashimoto:2020mrx} relying on symmetry arguments seems to fail.\footnote{Given a strip-like entangling region, the Hamiltonian associated with the bulk surface area involves a cyclic coordinate, thanks to which an explicit relation between the strip width and the maximum bulk reach of the minimal surface can be found. This then admits of explicit inversion of $f(z)$ in terms of area data. For spherical regions, similar symmetry arguments do not work: although in pure AdS a choice of coordinate system adapted to the dilatation symmetry of the AdS metric enables identification of a first integral of the Hamiltonian \cite{Jokela:2019ebz}, in our case this symmetry is in general spoiled by the presence of $f(z)$.} Therefore, instead of trying to solve the full problem right away we first consider the linearized problem. In the end, it turns out that the solution to the linearized problem can be used to construct the solution to the nonlinear problem with arbitrary precision.

\subsection{Scaling}\label{sec:scaling}

Consider now a perturbed metric 
\begin{equation}
    g_s = \frac{l^2}{z^2}\left( f_s(z) \dd z^2 + \euc\right) \ ,
\end{equation}
where $f_s(z) = 1 + s\, \delta f(z)$ and $\delta f$ is an unknown perturbation, $|s|\ll 1$, and $l$ is the radius of curvature. $e$ corresponds to the line element of other directions which do not play a key role in what follows. Minimal surfaces $\Sigma_s$ of the above metric have an infinite area. However, linearization of the area with respect to $s$ yields a finite quantity if we impose sufficiently strong decay conditions on the perturbation $\delta f$ near the conformal boundary. From a field theory perspective, this corresponds to focusing on deformations of the quantum state, {\emph{i.e.}}, no operators have sources that would trigger a renormalization group flow. Then we can pose the following inverse problem: \emph{Given linearized area data associated to hemispheres that foliate (at least an open subset of) the interior manifold, and its first and second radial variations, find $\delta f$.} A natural reference frame for studying this setup is the spherical coordinate system
\begin{align}
    x = \rho \sin\theta \cos\phi\ , \quad y = \rho\sin\theta\sin\phi \ , \quad z = \rho \cos\theta.
\end{align}
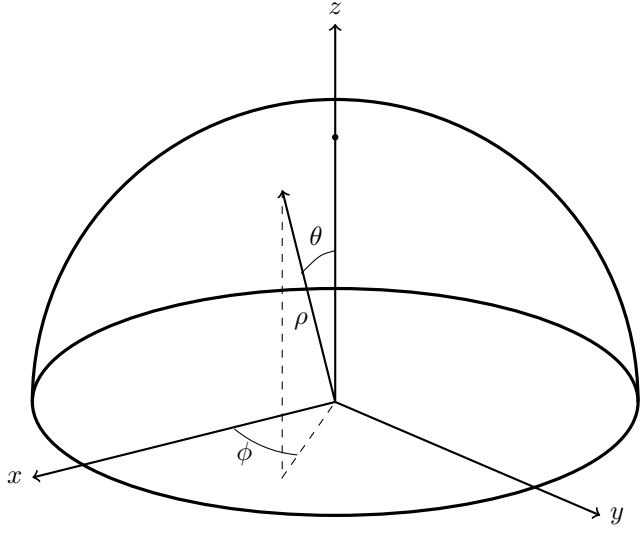
\begin{figure}
\begin{center}
\begin{tikzpicture}
    \draw[thick,->] (7,0) -- (7,5) node[anchor=south]{$z$};
    \draw[thick,->] (7,0) -- (3,-1) node[anchor=east]{$x$};
    \draw[thick,->] (7,0) -- (10.5,-1.5) node[anchor=west]{$y$};
    \draw[thick,->] (7,0) -- (6.3,2.8);
    \node[anchor=north] at (6.55,1.3) {$\rho$};
    \draw (7,2) .. controls (6.8,1.95) .. (6.55,1.7);
    \node[anchor=south] at (6.75,1.95) {$\theta$};
    \draw[dashed] (6.3,2.8) -- (6.3,-1) -- (7,0);
    \draw (6.5,-0.7) arc (-95:-130:1.5);
    \node[anchor=north] at (5.8,-0.4) {$\phi$};
    
    \filldraw  (7,3.5) circle (1pt);
    
    \draw[very thick] (7,0) ellipse (4 and 1.5);
    
    \draw[very thick] (11,0) arc (0:180:4);
\end{tikzpicture}
\end{center}
\caption{Our convention for the spherical coordinate system. Here $\theta=\pi/2$ corresponds to the conformal boundary.}
\end{figure}
Now we choose the radial coordinate $\rho$ instead of $z$ as the embedding function, which we parametrize in terms of the angles $\theta,\phi$. The minimal surface solution depends on $s$, which parametrizes the family of perturbed metrics, so we denote the solution by $\rho_s = \rho_s(\theta,\phi)$. Moreover, the area functional itself now explicitly depends on $s$ so we write it as
\begin{equation}
    A_s[\rho_s] = \int L_s[\rho_s,\na \rho_s] \ ,
\end{equation}
where $L_s$ is the associated Lagrangian that depends on $s$. Linearization of the area with respect to $s$ is
\begin{equation}\label{deltaA-general}
    \delta A (\rho_0) := \frac{d}{ds}\Big\rvert_{s=0} A_s [\rho_s] =\int \dot{L}_0 [\rho_0, \na \rho_0] \ ,
\end{equation}
where $\rho_0$ is the hemisphere solution satisfying 
\[
\rho_0(\theta=\frac \pi 2,\phi) = r > 0.
\]
Notice that after differentiating the integrand and applying the chain rule, only the $\dot{L}_0$ term is non-vanishing since $\rho_0$ is a minimizer. $\delta A$ is now a function that takes in a radius $r$ and outputs a real number.  An explicit calculation then gives that 
\begin{eqnarray}\label{deltaA'}
    \delta A (r) = \frac{l^2}{2}\int_0^{2\pi}\int_0^{\pi/2}\delta f(r \cos\theta) \tan^2\theta \sin\theta \,\dd\theta \dd\phi = \frac{\pi l^2}{r} \int_0^r \frac{r^2 - s^2}{s^2} \delta f(s)\dd s \ ,
\end{eqnarray}
where we performed a change of variables $s = r \cos \theta.$
We see that the integral is convergent if and only if $\delta f(z) \sim z^a$ for $a > 1$ as $z \to 0$.
Differentiating \label{deltaA} once with respect to $r$, we get
\begin{equation}
    \delta A'(r) = - \frac{\delta A(r)}{r} + 2\pi l^2 \int_0^r \frac{\delta f(s)}{s^2} \dd s  \ .
\end{equation}
Differentiating a second time gives
\begin{equation}
    \delta A''(r) = \frac{\delta A(r)}{r^2} - \frac{\delta A'(r)}{r} + 2\pi l^2 \frac{\delta f(r)}{r^2} \ ,
\end{equation}
from which we get the following inversion formula for $\delta f$
\bea
    \delta f(r) & = & \frac{1}{2\pi l^2}\left(
    - \delta A(r) + r\,\delta A'(r) +r^2 \delta A''(r)
    \right) \nn\\
    & = & \frac{4\G}{2\pi l^2}\left(
    - \delta S_{EE}(r) + r\,\delta S_{EE}'(r) +r^2 \delta S_{EE}''(r)
    \right)\ .\label{deltaf}
\eea
Hence, if the perturbed area data for hemispheres is known, corresponding to the entanglement entropy variations, the metric perturbation $\delta f$ can be uniquely recovered.  We will provide an explicit example below.

\paragraph{Generalization to higher dimensions.} Now we consider a perturbed metric on a time slice of $\text{AdS}_{n+1}$
\begin{equation}
    g = \frac{l^2}{z^2}\left( (1 + s\, \delta f(z))\dd z^2 + \euc\right)  \ ,
\end{equation}
where $\euc$ is the $(n-1)-$dimensional Euclidean metric. For simplicity, let us set $l=1$. Again, we transform into the spherical coordinate system $\rho, \theta, \phi^1,...,\phi^{n-2}$ 
\begin{align}
    x^1 &= \rho \sin\theta \sin\phi^1 \cdots \sin\phi^{n-2} \nn \\
    x^2 &= \rho \sin\theta\sin\phi^1 \cdots \sin\phi^{n-3}\cos\phi^{n-2} \nn \\ 
    &\vdots \nn \\ 
    x^{n-1} &= \rho \sin\theta \cos\phi^1 \nn \\
    z &= \rho \cos \theta \ ,
\end{align}
where the metric becomes
\begin{align}
    g = &\frac{1}{\rho^2\cos^2\theta}\left( (1 + s\, \delta f(z)\cos^2\theta)\dd \rho^2 - 2 s\, \delta f (z) \rho \sin\theta \cos\theta \dd\rho \dd\theta \right. \nn \\
    &\left. \hspace{1.5cm} \rho^2 (1 + s\, \delta f(z) \sin^2\theta) \dd\theta^2 + \rho^2 \sin^2\theta g_{S^{n-2}} \right) \ ,
\end{align}
where $g_{S^{n-2}}$ is the unit $S^{n-2}$ metric. For a fixed $s$, parametrize the minimal hypersurface $\Sigma_s$ by $\rho_s = \rho_s(\theta,\phi^i)$. The induced metric $g_{\Sigma_s}$ on the hypersurface looks very complicated but since $\rho_0$ corresponds to the $(n-1)$-hemisphere solution with $\pt_\theta \rho_0 = \pt_{\phi^i} \rho_0 = 0$, the only nonzero component of the $s$-derivative of the induced metric is
\begin{equation}
    (\dot{g}_{\Sigma_0})_{\theta\theta} = \tan^2 \theta\, \delta f(\rho_0 \cos\theta) \ .
\end{equation}
From the determinant formula
\begin{equation}
    \det (g_{\Sigma_s}) = \eps^{a_1...a_{n-1}}(g_{\Sigma_s})_{\theta a_1}(g_{\Sigma_s})_{\phi^1 a_2}\cdots (g_{\Sigma_s})_{\phi^{n-2}a_{n-1}} \ ,
\end{equation}
where $\eps^{a_1...a_{n-1}}$ is the Levi--Civita symbol, we get
\begin{equation}
    \frac{d}{ds}\Big\rvert_{s=0} \det (g_{\Sigma_s}) = \tan^{2(n-1)}\theta \, \delta f (\rho_0 \cos\theta) \det (g_{S^{n-2}}) \ .
\end{equation}
Perturbation of the hypersurface volume then becomes
\begin{align}
    \delta A(\rho_0)\equiv \frac{d}{ds}\Big\rvert_{s=0} A_s(\rho_s) &= \int_{S^{n-1}_+} \frac{d}{ds}\Big\rvert_{s=0} \sqrt{\det (g_{\Sigma_s})} \nn \\
    &= \frac{1}{2} \left( \int_{S^{n-2}}\sqrt{\det (g_{S^{n-2}})}  \right)\left( \int_0^{\pi/2}\delta f(\rho_0 \cos\theta)\tan^{n-1}\theta \sin\theta \dd\theta \right) \nn \\
    &= \frac{1}{2}\text{Vol}(S^{n-2}) \int_0^1 \delta f(\rho_0 x) \frac{(1-x^2)^{\frac{n-1}{2}}}{x^{n-1}}\dd x \ ,
\end{align}
where $S^{n-1}_+$ denotes the northern hemisphere.
Notice that to have a finite quantity on the right hand side, we need to impose the decay condition $\delta f(z) \sim z^a$ with $a > n-2$ when approaching the boundary. Denoting simply $r = \rho_0$ and making the change of variables $y = r x$ in the integral, we have
\begin{equation}
    \delta A (r) = \frac{1}{2}\text{Vol}(S^{n-2}) r\pminus{1}\int_0^r \delta f(y) \frac{(r^2-y^2)^{\frac{n-1}{2}}}{y^{n-1}}\dd y \ .
\end{equation}
Differentiating with respect to $r$ gives
\begin{equation}
    \delta A' (r) = - \frac{\delta A(r)}{r} + \frac{1}{2}\text{Vol}(S^{n-2}) (n-1)\int_0^r \delta f(y) \frac{(r^2-y^2)^{\frac{n-3}{2}}}{y^{n-1}}\dd y \ ,
\end{equation}
which we write in an equivalent form
\begin{equation}
    \frac{1}{r}\frac{d}{dr}(r\, \delta A(r)) = \frac{n-1}{2}\text{Vol}(S^{n-2})\int_0^r \delta f(y) \frac{(r^2-y^2)^{\frac{n-3}{2}}}{y^{n-1}}\dd y \ .
\end{equation}
We can solve this for $\delta f$ by iterative differentiation. For convenience, first introduce a differential operator $D_r \equiv r\pminus{1}(d/dr)$. Then acting $\ceil{(n-3)/2}$ times with $D_r$ on both sides yields
\begin{equation}\label{even-and-odd}
    D_r^{\ceil{\frac{n-1}{2}}}(r \,\delta A(r)) = \frac{(n-1)!!}{2}\text{Vol}(S^{n-2}) \times \begin{dcases}
        \int_0^r \frac{\delta f(y) }{y^{n-1}}\dd y, \quad &n \geq 3 \, \, \text{odd} \\
        \int_0^r \frac{\delta f(y) }{y^{n-1}\sqrt{r^2 - y^2}}\dd y, \quad &n \geq 4 \, \, \text{even} \ .
    \end{dcases} 
\end{equation}
Suppose now that $n \geq 3$ is \textbf{odd}.
Acting one more time with $D_r$, we finally get
\begin{equation}
    D_r^{\frac{n+1}{2}}(r \,\delta A(r)) = \frac{(n-1)!!}{2}\text{Vol}(S^{n-2}) \frac{\delta f(r)}{r^n} \ .
\end{equation}
Thus, we obtain an inversion formula
\begin{equation}\label{eq: odd inversion}
    \delta f(r) = \frac{8\G r^n}{(n-1)!! \text{Vol}(S^{n-2})l^{n-1}}D_r^{\frac{n+1}{2}}(r \,\delta S_{EE}(r))\ , \quad n \text{ odd}\ ,
\end{equation}
where we reinstated the AdS radius $l$.
With $n = 3$ this yields the formula \eqref{deltaf} we obtained above as a special case. 

Suppose then that $n\geq 4$ is \textbf{even}. We then have an integral equation of type \cite[Eq. (1.65)]{shaposhnikova1975integral}
\begin{equation}
   \int_a^x \frac{u(y)}{\sqrt{x^2-y^2}}\dd y =  h(x) \ .
\end{equation}
To solve this, we first replace $x$ by $s$, then multiply by $2s(x^2 - s^2)^{-1/2}$ and integrate with respect to $s$
\begin{equation}
    \int_a^x \frac{2s}{\sqrt{x^2-s^2}} \int_a^s \frac{u(y)}{\sqrt{s^2 - y^2}} \dd y \dd s = \int_a^x \frac{2s h(s)}{\sqrt{x^2-s^2}}\dd s \ .
\end{equation}
Exchanging the order of integration and making a change of variables $s = \sqrt{y^2 +(x^2 -y^2)t}$, we get on the left hand side
\begin{align}
    \int_a^x \frac{2s}{\sqrt{x^2-s^2}} \int_a^s \frac{u(y)}{\sqrt{s^2 - y^2}} \dd y \,\dd s &= \int_a^x \left( \int_y^x \frac{2s}{\sqrt{x^2 - s^2}\sqrt{s^2 - y^2}} \dd s \right) u(y)\dd y \nn \\
    &= \int_a^x \left( \int_0^1 \frac{1}{\sqrt{1-t}\sqrt{t}} \dd t\right) u(y)\dd y \nn \\
    &= \text{B}(1/2,1/2) \int_a^x u(y)\dd y \ ,
\end{align}
where $\text{B}$ is the beta function whose value at the given point is $\text{B}(1/2,1/2) = \pi$.
This then yields the solution
\begin{equation}
    u(x) = \frac{2}{\pi} \frac{d}{dx} \int_a ^x \frac{s h(s)}{\sqrt{x^2-s^2}}\dd s\ .
\end{equation}
Applying this formula to the even case in \nr{even-and-odd}, we obtain
\begin{equation}
    \frac{(n-1)!!}{2}\text{Vol}(S^{n-2}) \frac{\delta f(r)}{r^{n-1}} = \frac{2}{\pi} \frac{d}{dr} \int_0 ^r \frac{s D_s^{\ceil{\frac{n-1}{2}}}( s\, \delta A(s))}{\sqrt{r^2-s^2}}\dd s\ ,
\end{equation}
which yields the inversion formula
\begin{equation}\label{eq: even inversion}
    \delta f (r) = \frac{16\G r^{n-1}}{(n-1)!!\pi \text{Vol}(S^{n-2})l^n} \frac{d}{dr} \int_0 ^r \frac{s D_s^{\ceil{\frac{n-1}{2}}}(s\, \delta S_{EE}(s))}{\sqrt{r^2-s^2}}\dd s \ , \quad n \text{ even}\ ,
\end{equation}
where we again restored the AdS radius.

\paragraph{Example: AdS black hole.} To demonstrate our new techniques, we show how to recover the AdS-Schwarzschild geometry to linear order from boundary data. We focus here on the $n=3$ case and write the asymptotically AdS$_4$ metric in the gauge
\begin{equation}\label{eq:adsbh metric_intro}
    g = \frac{l^2}{z^2}\left(
    - f(z)\pminus{1} \dd t^2 + f(z)\dd z^2 + \dd x^2 + \dd y^2
    \right) \ .
\end{equation}
As before, we are sticking to the field theory being homogeneous and isotropic. The induced metric on a time slice reads
\begin{equation}\label{BH-slice}
    g_\rmi{ind} = \frac{l^2}{z^2}\left(f(z)\dd z^2 + \dd x^2 + \dd y^2
    \right)\ ,
\end{equation}
and our aim is to recover $f(z)$ at the perturbative level.

Our boundary data consists of temperature perturbation of entanglement entropy associated with a disk of radius $R$, which is given by (see Appendix~\ref{app:syntheticdata})
\begin{equation}\label{eq:T0data}
    \delta S_{EE}(R) = \frac{4\pi^4 l^2 R^3}{27 \G} \ .
\end{equation}
By the RT formula, this is dual to perturbation of the area of a minimal surface anchored on the boundary of the disk on the boundary.
The inversion formula \eqref{deltaf} then guarantees that the bulk geometry can be recovered from the data to linear order; the formula yields in this case
\begin{equation}\label{eq:deltafBH}
    \delta f(R) = \frac{1}{2\pi l^2} 4 \G \left(
    - \delta S_{EE}(R) + R\,\delta S_{EE}'(R) +R^2 \delta S_{EE}''(R)
    \right) = \frac{64 \pi^3 R^3}{27} \ .
\end{equation}
Identifying a parameter $s = (RT)^3$, we get
\be\label{eq: adsbh metric}
    f(z)  =  1+s \delta f +\ldots =  1 + \frac{64\pi^3 T^3}{27}z^3 + \ldots \, .
\ee
On the other hand, for the AdS-Schwarzschild metric the factor $f(z)$ in \eqref{BH-slice} can be expanded as a geometric series outside the black hole as
\begin{equation}\label{eq: adsbh blackening}
    f(z) = \frac{1}{1-\left(\frac{z}{z_H}\right)^3} = 1 + \left(\frac{z}{z_H}\right)^3 + \left(\frac{z}{z_H}\right)^6 + \ldots \, .
\end{equation}
Recalling the relation $T = 3/(4\pi z_H)$, we see that \eqref{eq: adsbh metric} indeed matches to first order with \eqref{eq: adsbh blackening}.

\section{Shape deformation}\label{sec:shapes}

Let us now look at a perturbed metric 
\begin{equation}\label{t-slice2}
    g_s = \frac{l^2}{z^2}\left( (1 + s\delta f(x,y,z)) \dd z^2 + \dd x^2 + \dd y^2\right)\ ,
\end{equation}
where the unknown perturbation $\delta f$ now depends on all the Cartesian coordinates $(x,y,z)$. It is again more convenient to work in spherical coordinates $(\rho,\theta,\phi)$ so let us rename our unknown function as
\begin{equation}
    \df(\rho,\theta,\phi) \equiv \delta f(\rho \sin\theta\cos\phi, \rho \sin\theta\sin\phi, \rho \cos\theta) = \delta f(x,y,z) \ .
\end{equation}
Since the perturbation has a more complicated dependence on the bulk coordinates, the simple analytic approach described in previous sections no longer works. Instead, we need to consider more general boundary curves for minimal surfaces and study the \emph{second linearization} of the perturbed minimal surface equation. Let $F = \epsilon_1 f_1 + \epsilon_2 f_2 + \epsilon_1 \epsilon_2 f_{12}: S^1 \to \reals$ be the boundary curve, and let $\rho_s^F$ be the minimal surface in $(M,g_s)$ with boundary curve given by $F$. First, the area perturbation reads 
\begin{equation}\label{Ldot-integral}
    \frac{d}{ds}\Big\rvert_{s=0} A_s (\rho_s^F) =\int \dot{L}_0 (\rho_0^F, \na \rho_0^F)\ ,
\end{equation}
where $\rho_0^F$ is the minimal surface in pure AdS given by the boundary curve $F$. As before, we assume that this is our regular data, which is the case only if $\delta \Tilde{f}$ satisfies certain fall-off conditions. To see this, write the above integral explicitly in spherical coordinates 
\begin{align}
    &\int \dot{L}_0 (\rho_0^F, \na \rho_0^F) \nn \\
    = &\int_0^{2\pi} \int_0^{\pi/2} \frac{\delta\Tilde{f}}{\cos^2\theta} \frac{ \left((\pt_\phi \rho_0^F)^2+\sin ^2\theta  \cos ^2\theta (\pt_\theta \rho_0^F )^2-2 \sin ^3\theta  \cos \theta  \rho_0^F \,\pt_\theta \rho_0^F +\sin ^4\theta  (\rho_0^F) ^2\right)}{2 \rho_0^F \sqrt{(\pt_\phi\rho_0^F  )^2+\sin ^2\theta (\pt_\theta\rho_0^F  )^2+\sin ^2\theta (\rho_0^F) ^2}} \dd\theta \dd\phi \ ,
\end{align}
where $\delta \Tilde{f} = \delta\Tilde{f}(\rho_0^F (\theta ,\phi ),\theta ,\phi )$ and $\rho_0^F = \rho_0^F(\theta,\phi)$. Recalling that
\begin{equation}
    \delta\Tilde{f}(\rho_0^F (\theta ,\phi ),\theta ,\phi ) = \delta f (\rho_0^F (\theta ,\phi )\sin\theta\cos\phi, \rho_0^F (\theta ,\phi )\sin\theta\sin\phi, \rho_0^F (\theta ,\phi ) \cos\theta)
\end{equation}
we observe that the integral is convergent only if $\delta f (x,y,z) = \mathcal{O}(z^2)$ as $z \to 0$, {\emph{i.e.}}, $\delta f(x,y,0 ) = \pt_z \delta f(x,y,0) = 0$ for all $x,y$. Then also $\pt_x \delta f (x,y,0) = \pt_y \delta f (x,y,0) = 0$ for all $x,y$.

The first linearization of \nr{Ldot-integral} with respect to boundary values is
\begin{equation}
    \frac{d}{d\eps_1}\frac{d}{ds}\Big\rvert_{s=0} A_s (\rho_s^F) = \int \left( \pt_\rho \dot{L}_0 (\rho_0^F, \na \rho_0^F) \frac{d}{d\eps_1}\rho_0^F + \pt_{\na\rho} \dot{L}_0 (\rho_0^F, \na \rho_0^F) \cdot \na\frac{d}{d\eps_1}\rho_0^F \right) \ .
\end{equation}
Denoting $v_i = \frac{d}{d\eps_i} \rho_0^F \rvert_{\eps_i=0}, i=1,2,$ and $ w = \frac{d}{d\eps_2}\frac{d}{d\eps_1}\rho_0^F\rvert_{\eps_1=\eps_2=0}$, the second linearization is
\begin{align}
    \frac{d}{d\eps_2}\frac{d}{d\eps_1}\Big\rvert_{\eps_1=\eps_2=0}\frac{d}{ds}\Big\rvert_{s=0} A_s (\rho_s^F) = \int \left( \right.
    &\left.\pt_\rho^2 \dot{L}_0 (\rho_0, \na \rho_0) v_1 v_2 + \pt_{\pt_i \rho}\pt_{\pt_j \rho}\dot{L}_0 (\rho_0, \na \rho_0) \pt_i v_1 \pt_j v_2 \right. \nn \\
    & \left.+\pt_\rho \pt_{\pt_i\rho} \dot{L}_0 (\rho_0, \na \rho_0) ( v_1 \pt_i v_2 + v_2 \pt_i v_1) \right.\nn \\
    & \left.+\pt_\rho \dot{L}_0 (\rho_0, \na \rho_0) w + \pt_{\pt_i\rho} \dot{L}_0 (\rho_0, \na \rho_0) \pt_i w \right)\ ,\label{integral-id}
\end{align}
where $\rho_0$ is the hemisphere solution; then, $\na\rho_0 = (\pt_\theta \rho_0, \pt_\phi \rho_0) = 0$. On the LHS we have our boundary data: second linearized area perturbations, {\emph{i.e.}}, second order small changes in area perturbations when the boundary disk is slightly deformed so that the deformation depends on two parameters. In the RHS integral we have terms containing derivatives of the Lagrangian and solutions to the first and second linearized minimal surface equation in pure AdS, $v_i$ and $w$ respectively. The functions $v_i$ can be chosen to be Hubeny's first order solutions
\begin{equation}
    u_l(\theta,\phi) = \tan^l(\theta/2)(1+l \cos\theta) e^{il\phi} \ ,
\end{equation}
whereas for $w$ we do not have explicit solutions ready at hand. Therefore we need to first solve the second linearized minimal surface equation, which reads (see Appendix \ref{app:linearization})
\begin{equation}\label{second-linearization}
    \sin^2\theta  \frac{\pt^2 w}{\pt \theta^2} + \frac{\pt^2 w}{\pt \phi^2} + \tan\theta \,(\sin^2 \theta + 1) \frac{\pt w}{\pt \theta} =  \frac{2}{\rho_0}\left( \sin^2\theta \frac{\pt u_k}{\pt\theta}\frac{\pt u_l}{\pt\theta} + \frac{\pt u_k}{\pt\phi}\frac{\pt u_l}{\pt\phi}\right) \ .
\end{equation}
The form of the LHS is identical to the first linearization of the minimal surface equation, but now the first order solutions appear on the RHS as a source so that the equation becomes inhomogeneous. Notice that if we have a solution $w$ to the inhomogeneous equation, then also $w + v$ solves the inhomogeneous equation when $v$ is a solution to the homogeneous equation, {\emph{i.e.}}, a solution to the first linearized minimal surface equation ({\emph{e.g.}}, one of Hubeny's solutions).

Remarkably, one can verify that given two first order solutions $u_k,u_l$ on the RHS as a source, one solution to the second linearized equation is
\begin{equation}\label{eq:2nd-order-sol}
    w = \rho_0\pminus{1} u_k u_l \ .
\end{equation}
Since the first order solutions satisfy the homogeneous equation, we are free to add any Hubeny's solution $u_j$ to the second order solution
\begin{equation}
    w = \rho_0\pminus{1} u_k u_l + u_j \ .
\end{equation}
For simplicity, we only consider second order solutions of the form \eqref{eq:2nd-order-sol}. With these choices of linearized solutions, the integral identity becomes
\begin{align}
    \frac{d}{d\eps_2}\frac{d}{d\eps_1}\Big\rvert_{\eps_1=\eps_2=0}\frac{d}{ds}\Big\rvert_{s=0} A_s (\rho_s^F) = \int \left( \right.
    &\left.(\pt_\rho^2 \dot{L}_0 (\rho_0, 0) + {\rho_0}\pminus{1}\pt_\rho \dot{L}_0 (\rho_0, 0)) u_k u_l \right. \nn \\
    & \left. + \pt_{\pt_i \rho}\pt_{\pt_j \rho}\dot{L}_0 (\rho_0,0) \pt_i u_k \pt_j u_l \right. \nn \\
    & \left.+(\pt_\rho \pt_{\pt_i\rho} \dot{L}_0 (\rho_0, 0) + {\rho_0}\pminus{1}\pt_{\pt_i\rho} \dot{L}_0) ( u_k \pt_i u_l + u_l \pt_i u_k) \right)\ .\label{int-id}
\end{align}
The above Lagrangian derivatives are, explicitly,
\begin{align}
    \pt_\rho \dot{L}_0 (\rho_0, 0 ) &= \frac{1}{2} \sin \theta  \tan ^2\theta  \pt_\rho \delta \Tilde{f}(\rho_0 ,\theta ,\phi ) \\
    \pt_\rho^2 \dot{L}_0 (\rho_0, 0 ) &= \frac{1}{2} \sin \theta  \tan ^2\theta  \pt_\rho^2 \delta \Tilde{f}(\rho_0 ,\theta ,\phi ) \\
    \pt_{\na \rho} \dot{L}_0 (\rho_0, 0 ) &= \left(-\frac{\sin \theta  \tan \theta  \delta \Tilde{f}(\rho_0 ,\theta ,\phi )}{\rho_0 },0\right) \\
    \pt_{\na \rho} \pt_\rho \dot{L}_0 (\rho_0, 0 ) &= \left(\frac{\sin \theta  \tan \theta  \left(\delta \Tilde{f}(\rho_0 ,\theta ,\phi )-\rho\,  \pt_\rho\delta \Tilde{f}(\rho_0 ,\theta ,\phi )\right)}{\rho_0 ^2},0\right) \\
    \left(\pt_{\pt_i \rho}\pt_{\pt_j\rho}\dot{L}_0 (\rho_0, 0 )\right)_{i,j\in\{\theta,\phi\}}&= \frac{\delta \Tilde{f}(\rho_0 ,\theta ,\phi )}{4 \rho_0 ^2 \cos^2\theta}
    \begin{pmatrix}
    (1+3 \cos 2 \theta ) \sin \theta  & 0 \\
    0 & (3+\cos 2 \theta ) \csc \theta    \\
    \end{pmatrix} \ .
\end{align}
Next, we expand $\delta\Tilde{f}$ in spherical harmonics as
\begin{equation}
    \delta \Tilde{f}(\rho ,\theta ,\phi ) = \sum_{l,m} a_{lm}(\rho) Y^m_l (\theta,\phi) \ ,
\end{equation}
where $a_{lm}=0$ whenever $l-m$ is even; this enforces the premise that $\delta \Tilde{f}$ vanishes on the boundary when $\theta \to \pi/2$ with $\rho > 0$ fixed. On the other hand, the series has to vanish when $\rho \to 0$ with $\theta \in (0,\pi/2)$ fixed, from which (using Dominated Convergence) we have that $a_{lm}(0) = 0$ for all $l,m$. Moreover, since we found above that $\na \delta f (x,y,0) = 0$, we have
\begin{align}
    \pt_\rho \delta \Tilde{f}(\rho, \theta, \phi)  = \pt_\rho \delta f (x,y,z) = (\sin\theta\cos\phi, \sin\theta\sin\phi, \cos\theta) \cdot \na \delta f (x,y,z) \to 0 \ ,
\end{align}
as $\rho \to 0$ with $\theta,\phi$ fixed. This implies
\begin{equation}
    \lim_{\rho \to 0} \sum_{l,m} a_{lm}'(\rho) Y^m_l (\theta,\phi) = 0  \ , \quad \forall \,\theta,\phi \ ,
\end{equation}
which gives the initial condition $a_{lm}'(0) = 0$ for all $l,m$. We will need the initial data $a_{lm}(0) = a_{lm}'(0) = 0$ for an ODE system later on.

Since the normal derivative of $\delta \Tilde{f}$ vanishes on the boundary, we get
\begin{equation}\label{normal-derivative}
    0 = \frac{d}{dz}\Big\rvert_{z=0}\sum_{l,m}a_{lm}(\rho) Y^m_l (\theta,\phi) = \sum_{l,m}\left( a_{lm}'(\rho)\cos\theta Y^m_l (\theta,\phi) -\rho\pminus{1} a_{lm}(\rho)\sin\theta \pt_\theta Y^m_l (\theta,\phi)
    \right)\big\rvert_{z=0} \ ,
\end{equation}
where we used the chain rule and the Jacobian matrix of the coordinate transformation $(\rho,\theta,\phi) \to (x,y,z)$. Taking $\theta\to\pi/2$ with other coordinates fixed, we get that 
\begin{equation}
    \sum_{l,m} a_{lm}(\rho) \pt_\theta Y^m_l (\pi/2,\phi) = 0\ ,\quad \forall \rho \geq 0, \, \forall \phi \in [0,2\pi) \ .
\end{equation}
Multiplying by $e^{-i n \phi}$ and integrating over $[0,2\pi)$ with respect to $\phi$, this gives
\begin{equation}\label{deriv-constraint}
    \sum_{l=|n|}^\infty a_{ln}(\rho) \pt_\theta Y^n_l (\pi/2,0) = 0 \ ,\quad \forall \rho \geq 0 \ .
\end{equation}

We can approximate $\delta \Tilde{f}$ to an arbitrary precision by a finite sum,
\begin{equation}
    \delta \Tilde{f}(\rho ,\theta ,\phi ) \approx \sum_{l=0}^n \sum_{m=-l}^l a_{lm}(\rho) Y^m_l (\theta,\phi) \ ,
\end{equation}
where $n>0$ since the only constant mode satisfying the boundary conditions is zero.
Here we have $(n+1)^2$ unknown coefficients to begin with, but we get $(n+2)(n+1)/2$ constraints from the fact that $a_{lm} = 0$ for $l-m$ even. Moreover, for a finite sum \nr{deriv-constraint} becomes
\begin{equation}
    \sum_{l=|m|}^n a_{lm}(\rho) \pt_\theta Y^m_l (\pi/2,0) = 0\ ,\quad \forall \rho \geq 0\ , \, \forall m= -n,\ldots,n\ ,
\end{equation}
which amounts to $2n-1$ constraints for the remaining nonzero coefficients. All in all, we have $n(n+7)/2$ constraints, which leaves us with a total $(n^2 - 3n +2)/2$ unknown coefficients. This means that the problem of recovering the unknown coefficients is nontrivial only for $n\geq 3$.

\paragraph{Example case:} Suppose we set the series cutoff to $n=4$ and try to recover $\df$ approximated by
\begin{equation}
    \delta \Tilde{f}(\rho ,\theta ,\phi ) \approx \sum_{l=0}^4 \sum_{m=-l}^l a_{lm}(\rho) Y^m_l (\theta,\phi) \ .
\end{equation}
Constraints dictate that the right hand side sum only have three \emph{a priori} non-zero coefficients: $a_{1,0}, a_{2,\pm1}$. Denote by $\delta_{k,l} A (r)$ the second linearization of the area data corresponding to Hubeny's solutions $u_k, u_l$ in \eqref{int-id}. With suitable choices of $k,l$, the integral identity then yields a set of ODEs 
\begin{align}
    \delta_{-4,3} A(r)&=\frac{\sqrt{\frac{\pi }{30}} \left(r  \left(3618 a_{2,1}'(r )+319 r a_{2,1}''(r)\right)+24750 a_{2,1}(r)\right)}{12 r ^2} \nn \\
    \delta_{-4,4} A(r)&=-\frac{5 \sqrt{\frac{\pi }{3}} \left(r  \left(77 a_{1,0}'(r)+13 r a_{1,0}''(r)\right)+1248 a_{1,0}(r)\right)}{24 r ^2} \nn \\
    \delta_{-3,4} A(r)&=-\frac{\sqrt{\frac{\pi }{30}} \left(r  \left(162 a_{2,-1}'(r)+31 r a_{2,-1}''(r)\right)+1710 a_{2,-1}(r)\right)}{12 r ^2} \ .
\end{align}
Given the initial conditions $a_{lm}(0)=a'_{lm}(0)=0$ and the area data, we have a unique solution $a_{lm}$, which can be obtained analytically in terms of $\delta_{k,l} A$.

\section{Progress in nonlinear problem}\label{sec:nonlinear}

 Next, we consider the problem of reconstructing the full metric function $f$ in~\eqref{eq:metric_intro}, 
rather than just its perturbation $\delta f$. We assume the knowledge of the area data $A_f$ for minimal surfaces anchored at the boundary $z=0$ 
in the geometry determined by $f$. Under the assumption that we are close to pure AdS, $|f(z)-1|\ll 1$, for all $z$, 
we develop a reconstruction method for $f$. 

The key to our approach lies in the explicit representation for $\delta f$ in term of the linearized area functional $\delta A$, using 
the integral formula~\eqref{deltaf}. Notably, our method requires only the boundary 
area data $A_f$ as input. All other quantities appearing in the reconstruction can be computed 
directly from the bulk metric~\eqref{t-slice}.

\subsection{Abstract setting}
We now present the method in abstract terms, generalizing the setting slightly. 
The area data can be expressed as a nonlinear operator $A$ acting on functions $f=f(z)$ defined as in \eqref{eq:area-function} by 
\begin{equation}\label{eq:area_data_for_general_f}
 A[f](r):= l^{n-1} \int_{B(0,r)} \frac{\sqrt{1 + f(z_f(x; r))\abs{\na z_f(x; r)}^2_{\euc}}}{z_f(x; r)^{n-1}}\sqrt{\det(\euc)}\dd x^1 \wedge\dots\wedge \dd x^{n-1} \ ,
 \end{equation}
    where $z_f(\ccdot; r)$ is the minimal surface solution for the metric
    \begin{equation}\label{metric_nonlinear_sec}
 g(z,x) = \frac{l^2}{z^2}\left( f(z) \, \dd z^2 + e \right)
\end{equation}
    with boundary condition $|x|^2 = r^2$ on the $(x^1,\dots, x^{n-1})$ plane, and $B(0,r)$ is a ball of radius $r$ in $\mathbb R^{n-1}$. Thus, $A$ maps a given function $f$ and radius $r$ to the area of the minimal surface anchored at infinity as a sphere of radius $r$. 
    
The linearization of $A$ at $f=f_0$ is the operator $DA|_{f=f_0}$ defined as usual by
\[
DA|_{f=f_0} [w]:= \frac{d}{dt}\Big|_{t=0} A_{f_0+t w} \ . 
\]
We will consider the operator $DA|_{f=f_0}$ acting on various functions, but in the case $w=\delta f$, we write  
\[
 \delta A=DA|_{f=f_0} [\delta f]
\]
in accordance with our earlier notation. 
For brevity, we also denote
\[
 A_f=A[f] \ . 
\]

With these definitions we are now  ready to describe a method to recover an unknown function $f = f(z)$ in the metric \eqref{metric_nonlinear_sec} from its area data $A_f$. The approach assumes:
\begin{enumerate}
    \item[(1)] $f$ is a sufficiently small perturbation of a known function $f_0$.
    \item[(2)]\label{assum2} We have an explicit expression for the ``inverse" $(DA|_{f=f_0})^{-1}$ of the linearized operator in the sense that $(DA|_{f=f_0})^{-1} (DA|_{f=f_0}) = Id$. Though this operator need not be a two-sided inverse.
\end{enumerate}  
We note that the latter condition in the special case when $f_0 = 1$ is satisfied. In this case, the inverse operator $(DA|_{f=f_0})^{-1}$ acting on any $u = u(r)$ is given explicitly by 
 \begin{equation}\label{eq:inversion_intro_nonlinear_sec}
\Big((DA\mid_{f=f_0})^{-1}[u]\Big)(z) := \frac{1}{2\pi}\left( -u(z) + z\, u'(z) + z^2 u''(z) \right) \ ,
\end{equation}
as we have shown when $n=3$, see \eqref{deltaf} in  Section \ref{sec:scaling}. The above means that if $u = DA\mid_{f =1}[\delta f]$, then $\delta f(z)$ is given by \eqref{eq:inversion_intro_nonlinear_sec}:
\begin{eqnarray}
    \label{eq: verification_of_left_inverse}
    \delta f(z) = \frac{1}{2\pi}\left(-u(z) + zu'(z) + z^2 u''(z)\right) \ .
\end{eqnarray}

The penultimate equality follows from \eqref{deltaf}. For dimensions $n > 3$, similar inversion formulas are given in \eqref{eq: even inversion} and \eqref{eq: odd inversion}. We see that assumption (2) above is satisfied when $f_0= 1$.

For general $f_0$ (and not necessarily $f_0=1$) we assume the condition (2) above. Assume also that we are given the area data $A_f$ that corresponds to an unknown function $f$ assumed to be sufficiently close to $f_0$, 
\[
 |f-f_0|=\delta \ ,
 \]
for some $\delta \ll 1$.  We define the operator $F$ acting on functions $v=v(z)$ by 
\begin{eqnarray}
    \label{eq: def F operator}
F[v] := v-\left(DA\mid_{f =f_0}\right)^{-1}\left(A[v] - A_{ f}\right) \ .
\end{eqnarray}
We stress that to compute $F$ for given $v$, we only need to known the quantity $A_f$. The other quantities $\left(DA\mid_{f =f_0}\right)^{-1}$ and $A[\ccdot]$ can be computed using the form \eqref{ads-metric} of the metric.  Especially, for any $v$, $A[v]$ is obtained by solving the minimal surface equation with $v$ in place of $f$ in the metric \eqref{ads-metric} and substituting into \eqref{eq:area_data_for_general_f}. 

Our goal is to find a fixed point $v^*$ for $F$ satisfying $|v^*-f_0|<\delta$. That $v^*$ is a fixed point means
\[
 F[v^*]=v^*\ .
\]
We now demonstrate that such a fixed point must coincide with the unknown function $f$ we seek to recover from the area data $A_f$.
To this end, assume then that $v^*$ is a fixed point with $|v^*-f_0|<\delta$. 
It follows that
\begin{equation}\label{eq:fixed_point_cond}
 0=\left(DA\mid_{f =f_0}\right)^{-1}\left(A[v^*] - A_{ f}\right) \ .
\end{equation}
We Taylor expand at $v^*$ as 
\begin{multline}\label{eq: expand around f}
A[v^*] -A_{f}= DA\mid_{f= v^*}(v^*-f) + {\cal{O}}(|v^* -  f|^2) \\
= DA\mid_{f= f_0}(v^*-f) + \left(DA\mid_{f = v^*}- DA\mid_{f=f_0}\right)(v^*-f)+ {\cal{O}}(|v^* -  f|^2) \ ,
\end{multline}
where $|\ccdot|$ is some suitable norm we do not specify and the implied constant in the $O$-notation  bounded since we assumed $|v^*-f|\leq |v^*-f_0| +|f_0-f|\leq 2\delta$. 
Substituting the expansion \eqref{eq: expand around f} into \eqref{eq:fixed_point_cond} shows that 
\begin{multline}\label{eq:expansion2}
 0=\left(DA\mid_{f =f_0}\right)^{-1}\left(DA\mid_{f= f_0}(v^*-f) + \left(DA\mid_{f = v^*}- DA\mid_{f=f_0}\right)(v^*-f)+ {\cal{O}}(|v^* -  f|^2)\right) \\
 =v^*-f+\left(DA\mid_{f =f_0}\right)^{-1}\left(DA\mid_{f = v^*}- DA\mid_{f=f_0}\right)(v^*-f)+{\cal{O}}(|v^* -  f|^2) \ .
\end{multline}

We may also Taylor expand at $f_0$ as 
\[
 DA\mid_{f = v^*}- DA\mid_{f=f_0}={\cal{O}}(|v^* -  f_0|) \ ,
\]
with the implied constant in the $O$-notation is bounded due to $|v^*-f_0|<\delta$. 
Substituting this into \eqref{eq:expansion2} gives
\[
 v^*-f={\cal{O}}(|v^* -  f||v^* -  f_0|)+{\cal{O}}(|v^* -  f|^2) \ .
\]
Since $|v^*-f_0|,|v^*-f| <\delta$, this implies  
\[
 |v^* -  f|\leq \delta K |v^* -  f|
\]
for some $K>0$  independent of $\delta$. 
From this it follows that if $\delta$ was small enough, it must be
\[
 v^*=f \ . 
\]
(Otherwise one may divide by $|v^* -  f|$ to arrive at a contradiction.) 
Thus, we have shown that when $|f-f_0| = \delta$ with $\delta$ sufficiently small, any fixed point $v^*$ of $F$ with $|v^*-f_0|<\delta$ must coincide with the the function $f$ that we wanted to recover.

We next argue that the iteration defined by
\begin{eqnarray}\label{eq: iteration}
 v_{j+1}:=F[v_j] \ , \quad v_0=f_0
\end{eqnarray}
converges to a unique fixed point of $F$ satisfying $|v^*-f_0|<\delta$ provided that $\delta$ was small enough. Thus, by what we argued above, the iteration will then converge to $v^*=f$.  Note that 
if $v_1, v_2$ are two functions with $|v_1-f_0|,|v_1-f_0|<\delta$, we can take the expansion \eqref{eq: expand around f} with $v_1$ and $v_2$ in place of $v^*$ and $f$ respectively then hit both sides of the equation with $\left(DA\mid_{f=f_0}\right)^{-1}$ to get 
\[
 F[v_1] - F[v_2] = DF\mid_{f=v_2} (v_1-v_2) + {\cal{O}}(|v_1-v_2|^2) \ ,
\]
where the implied constant in the $O$-notation is bounded due to $|v_1-v_2|\leq 2\delta$. 
Note that
\begin{eqnarray*}
DF\mid_{v_2}=\left(DA\mid_{f=f_0}\right)^{-1}\left(DA\mid_{f = f_0} - DA\mid_{v_2}\right) \ ,
\end{eqnarray*}
which implies that 
\[
 DF\mid_{v_2}(v_1-v_2)={\cal{O}}(|v_1-v_2||f_0-v_2|)\ .
\]
Consequently, if $\delta$ was small enough (forcing $|f_0-v_2|$ and $|v_1-v_2|$ to be small), the operator  $F$ is a contraction in the sense that
\begin{eqnarray}\label{eq: contraction}|F[v_1] - F[v_2]| \leq C|v_1-v_2|
\end{eqnarray}
for some constant $C<1$ independent of $\delta$. 

By the contraction property \eqref{eq: contraction}, we have
\[
|v_{j+1} - v_j| = |F[v_j] - F[v_{j-1}]| \leq C |v_j - v_{j-1}| \ .
\]
Iterating this estimate gives
\[
|v_{j+1} - v_j| \leq C^j |v_1 - f_0| \ .
\]
Consequently, for any \( m > l \), we have by the triangle inequality:
\[
|v_m - v_l| \leq \sum_{k=l}^{m-1} |v_{k+1} - v_k| \leq |v_1 - f_0| \sum_{k=l}^{m-1} C^k \ .
\]
Summing the geometric series on the right gives
\[
|v_m - v_l| \leq \frac{C^l}{1 - C} |v_1 - f_0|\ .
\]
Since $C<1$, we thus have demonstrated the iteration $v_{j+1}=F[v_j]$ converges to a fixed point. Moreover, the convergence is geometrical. We also remark that the fixed point is unique and at a distance less than $\delta$ from $f_0$, which follow directly from the fact that $F$ was a contraction. In summary
\begin{equation}\label{eq:iterative-limit}
 \lim_{j\to \infty}v_j=v^*=f \ .
\end{equation}

The presentation here is, of course, not mathematically rigorous. We have not specified which norm $\|\cdot\|$ we are using on function spaces and whether the operators $A$, $DA|_{f=v^*}$ and $\left(DA\mid_{f=f_0}\right)^{-1}$ etc. behave well on these function spaces. A rigorous mathematical study of the limiting behavior of the iterative procedure \eqref{eq: iteration} is beyond the scope of this current article. We remark that analogous and mathematically precise arguments appear for example in \cite{stefanov2009linearizing}.

\subsection{Implementing the iterative algorithm: preliminary observations}

For concreteness, let us spell out the first few steps in the iterative procedure. Denote the points in the iteration by $f_i$. Our zeroth approximation to the unknown function $f$ is given by $f_0 \equiv 1$, which corresponds to the pure AdS. The next steps are
\begin{align}\label{eq: series rep}
    f_1 &= F(f_0) = f_0 - (D A\rvert_{f_0})\pminus{1}(A_{f_0} - A_f) \\
    f_2 &= F(f_1) = f_0 - (D A\rvert_{f_0})\pminus{1}(A_{f_0} - A_f) - (D A\rvert_{f_0})\pminus{1}(A_{f_1} - A_f) \\
    &\vdots \nn
\end{align}
Thus, we are in effect constructing a series representation for the unknown $f$. Notice the analogy to Newton's method in the above algorithm; where the analogy fails is that the inverse of the `derivative' is here always evaluated at the same point. Remarkably, for any function $\delta A$ we have an explicit expression for $\delta f= D^{-1} A\mid_{f_0} \delta A$ given by \eqref{eq: even inversion} (in even dimension) and \eqref{eq: odd inversion} (in odd dimension). This is all we need in terms of metric perturbation theory; we don't need explicit formulas for higher order perturbations $\delta^2 f, \delta^3 f$ etc.

We remark here that the series representation \eqref{eq: series rep} applies for reconstructing metrics which are near the AdS-Schwarzschild metric given in \eqref{eq: adsbh metric}, at least in the low temperature regime when $z_H\gg 1$. Indeed, since
$$f_{\rm AdS-BH}(z) = \frac{1}{1-(z/z_H)^3}$$
we see that $f_{\rm AdS-BH} \approx 1$ when $z_H\gg 1$. So any metric $\hat f$ which are near $f_{\rm AdS-BH}$ when $z_H\gg 1$ are also close to $1$ and therefore can be represented via \eqref{eq: series rep}.

Even though the area quantities above are in fact just the `finite parts' thereof, in practice we take $A_{f_i}$ and $A_f$ to be the non-regularized areas and write $A_\rmi{data} = A_{f_0}- A_f$, which is actually our regular data. Then we can augment the above procedure by writing $A_{f_i} - A_f = (A_{f_i} - A_{f_0}) + A_\rmi{data}$ so that at each iteration we regularize the new area function $A_{f_i}$ against the corresponding quantity in pure AdS (by combining two divergent integrals under a single integral sign so that the divergences in the integrands cancel out).

Numerical implementation of the algorithm may be rather challenging. For this, we first need to numerically solve the full minimal surface equation in the metric involving $f_i$ for a large family of boundary disk radii. Then we evaluate numerically the integral $A_{f_i} - A_{f_0}$ where we plug in the numerical approximation for the embedding function of the minimal surface corresponding to $f_i$, and the hemisphere solution, and we do this for a number of radii. Finally, we need a finite difference approximation for the first and second derivatives of $(A_{f_i} - A_{f_0}) + A_\rmi{data}$ to compute the next function $f_{i+1}$ using the explicit inversion formula. Controlling the numerical error through all these stages might require some care. The computation is probably going to be quite costly unless we for some reason have very fast convergence towards the sought-after function.

Our iterative scheme is reminiscent of the perturbative reconstruction method presented in \cite[Sec. 6.1]{Bilson:2010ff}. One notable difference, however, is that our algorithm is not based on a Taylor expansion of the unknown function. Our algorithm is global in the sense that each new step integrates over all the values of earlier iteration.  As it is well-known that Taylor expansion is highly unstable and sensitive to numerical errors, it is possible that our method leads to better stability and convergence in the reconstruction.

\section{Discussion}\label{sec:discussion}

Our analysis has focused on reconstructing aspects of the bulk geometry from boundary data, but several subtleties remain. One key omission in our discussion is the divergence structure of the entanglement entropy and how it behaves under changes in the cutoff. The presence of universal terms, particularly in scenarios where the boundary deformation includes cusps, introduces additional log-divergences, whose coefficients are often of physical significance. A more detailed treatment of these divergences could provide further insight into the universality of our reconstruction procedure.  

Beyond leading-order corrections, a natural extension of our approach is to reconstruct the full function $f(z) \approx 1 $ rather than just its perturbation $\delta f(z)$. The limit of the iterative formula (\ref{eq:iterative-limit}), which is just the full $f(z)$, suggests that, in principle, the entire black hole metric \eqref{eq:adsbh metric_intro} could be reconstructed, particularly in the regime where $z_H \gg 1$ (or equivalently, at low temperatures). This would provide a more complete picture of the emergent bulk spacetime from boundary data.  

Several directions remain for future work. One important path is to understand how uncertainties in the boundary data propagate to the reconstructed $\delta f$~\cite{Jokela:2020auu}, which is essential for practical applications. Another promising direction is the extension of our method to time-dependent setups, such as holographic quenches~\cite{Nozaki:2013wia}, where the HRT prescription  replaces the RT formula. This would allow us to explore the dynamical aspects of bulk reconstruction and their implications for nonequilibrium holography.  

Finally, while our work represents an early step in understanding bulk reconstruction from a mathematical perspective, it opens the door to further refinements and generalizations. By incorporating more sophisticated mathematical tools and addressing these open questions, we hope to move closer to a more complete picture of how spacetime emerges from field theory data.


\begin{acknowledgments}
We thank Jani Kastikainen and Esko Keski-Vakkuri for useful discussions.  N.~J. has been supported in part by the Research Council of Finland grant no.~13545331. T.~L. was partly supported by the Academy of Finland (Centre of Excellence in Inverse Modelling and Imaging and FAME Flagship, grant numbers 312121 and 359208). M.~S. was supported by the European Research Council of the European Union, grant 101086697 (LoCal),
and the Research Council of Finland, grants 347715,
353096 (Centre of Excellence of Inverse Modelling and Imaging)
and 359182 (Flagship of Advanced Mathematics for Sensing Imaging and Modelling).  L.~T. was partially supported by Australian Research Council DP190103451 and DP220101808. Views and opinions expressed are those of the authors only and do not necessarily reflect those of the European Union or the other funding
organizations.
\end{acknowledgments}

\appendix

\section{Synthetic data}\label{app:syntheticdata}

In this appendix we will generate the boundary data that can be used to infer the corresponding metric deformation in the bulk spacetime. We will focus on three-dimensional CFTs and obtain the change in the entanglement entropy for disks of radii $R$ with an infinitesimal temperature deformation about zero temperature.

Consider the minimal surface in the metric \eqref{BH-slice} anchored on a disk of radius $R$ on the boundary. In cylindrical coordinates $(u,\phi,z)$ related to the Cartesian coordinates by
\begin{equation}
    x = u \cos\phi\ , \quad y = u \sin\phi\ , \quad z = z\ ,
\end{equation}
the minimal surface area function $A(R)$ reads
\begin{equation}\label{eq:ARBHintegral}
    A(R) = 2 \pi l^2 \int_0^R \sqrt{\frac{1 + z'(u)^2 - (z(u)/z_H)^3}{1 - (z(u)/z_H)^3}}\frac{u}{z(u)^2}\dd u\ ,
\end{equation}
where $z(u)$ is the minimal surface embedding with $z(u=R) = 0$. By the RT formula, this corresponds to the entanglement entropy of a disk of radius $R$ in the boundary CFT at finite temperature $T = 3/(4\pi z_H)$. Linearizing the entanglement entropy around $T=0$ then corresponds to linearizing the minimal surface embedding $z(u)$ and then the area function around $z_H= \infty$, \emph{i.e.}, around pure AdS:
\bea
    A(R) & = & 2 \pi l^2 \left( \int_0^R \frac{\sqrt{1 + z_0'(u)^2}}{z_0(u)^2}u\dd u + \frac{1}{2 z_H^3}\int_0^R \frac{z_0(u)z_0'(u)^2}{\sqrt{1 + z_0'(u)^2}} u \dd u + \mathcal{O}(z_H\pminus{6}) \right) \\ \label{eq:ARBHintegral2}
    & = & 2 \pi l^2 \left( \int_0^1 \frac{\sqrt{1 + z_0'(x)^2}}{z_0(x)^2}x\dd x + \frac{32\pi^3(RT)^3}{27}\int_0^1 \frac{z_0(x)z_0'(x)^2}{\sqrt{1 + z_0'(x)^2}} x \dd x + \mathcal{O}((RT)^{6}) \right) \ ,
\eea
where we revealed the dimensions by substitution $u=Rx$, $z_0(u)=Rz_0(x)$, and where $z_0(x) = \sqrt{1 - x^2}$ is the minimal surface embedding $\Sigma$ in pure AdS -- just the usual hemisphere embedding anchored on the boundary entangling surface $\partial\Sigma$, a disk with radius $R$. The first term corresponds to the standard UV-divergent entanglement entropy of a disk of radius $R$ at $T=0$, which plays no role in our analysis.
The rest of the terms are dual to the entanglement entropy density, which are obtained upon a finite deformation by temperature \cite{Gushterov:2017vnr},
\be
 \sigma = \frac{S_{EE}-S_{EE}|_{T=0}}{{\rm{Vol}}{(\partial\Sigma)}} \ .
\ee
Here $\sigma$ is a UV-finite quantity by construction and hence will be a good candidate for the data too.

In what follows, we focus on only the second term, giving rise to the entanglement entropy deformation strictly at $T=0$. By making contact with the notation in the main text,
\bea
 A(R) & = & A(R)|_{s=0} + s\frac{d}{ds}A_s|_{s=0} + \ldots \\
 & = & A(R)|_{T=0} + T^3\frac{d}{dT^3}A_T|_{T=0} + \ldots \\
& = & A(R)|_{T=0} + T^3\delta A(R) +\ldots \ ,
\eea
where we identified $s\propto (RT)^3$. It is useful to view the perturbation as obtained by keeping the radius of the disk $R$ fixed and varying the temperature, but since this is a perturbation in the underlying CFT, it is equivalent to keep $T=0$ and vary the radius instead. In other words, the entangling surface $\cal A$ is kept intact, in this case the disk of area $\pi R^2$, but the area of the {\emph{dual}} hanging surface changes upon deformation by the temperature.
We further note that the perturbation arises at the third order power in $T$ as we are not deforming the Hamiltonian of the theory but the state.

An explicit calculation of the integral in the second term in \eqref{eq:ARBHintegral2} gives 
\begin{equation}\label{eq:T0data2}
    \delta A(R) = 4\G\delta S_{EE}(R) = \frac{16\pi^4 l^2 R^3}{27} \ .
\end{equation}
This is our `synthetic' boundary data: it was obtained from a bulk computation but now we pretend that it actually came from the boundary CFT, due to lack of existing results produced directly in the field theory.

\section{Linearizations of the minimal surface equation}\label{app:linearization}

As in \cite{Hubeny:2012ry}, we want to linearize the minimal surface equation in pure AdS$_4$ around the hemisphere solution so we set $f=1$ in \eqref{eq:nonlinear-minsurf}. As it is easier to work in coordinates that are natural for this problem, we transform the minimal surface equation \eqref{eq:nonlinear-minsurf} into spherical coordinates and obtain the same equation as Eq. (3.46) in \cite{Hubeny:2012ry}. To linearize the equation to second order with respect to two parameters $\eps_1,\eps_2$, we expand the solution $\rho(\theta,\phi)$ as
\begin{equation}\label{eq: eps expansion}
    \rho(\theta,\phi) = \rho_0 + \eps_1 u_1(\theta,\phi) + \eps_2 u_2(\theta,\phi) + \frac{1}{2}(\eps_1^2 v_1(\theta,\phi) + \eps_2^2 v_2(\theta,\phi) + 2\eps_1\eps_2 w(\theta,\phi)) + \mathcal{O}(\eps^3) \ ,
\end{equation}
and plug it into the minimal surface equation. The first linearized minimal surface equation becomes
\begin{equation}
    \sin^2 \theta \frac{\pt^2 u}{\pt \theta^2} + \frac{\pt^2 u}{\pt \phi^2} + \tan\theta \,(\sin^2 \theta + 1) \frac{\pt u}{\pt \theta} = 0 \ .
\end{equation}
The equation is the same for both $u_1$ and $u_2$ so we dropped the subscript for simplicity.
Using a separation of variables ansatz $u(\theta,\phi) = \Theta(\theta)\Phi(\phi)$, we obtain
\begin{equation}
    \sin^2 \theta \,\Theta''(\theta)\Phi(\phi) + \Theta(\theta)\Phi''(\phi) + \tan\theta \,(\sin^2 \theta + 1) \Theta'(\theta)\Phi(\phi) = 0\ ,
\end{equation}
which we can write as
\begin{equation}
    \sin^2 \theta \,\frac{\Theta''(\theta)}{\Theta(\theta)}+ \tan\theta \,(\sin^2 \theta + 1) \frac{\Theta'(\theta)}{\Theta(\theta)} + \frac{\Phi''(\phi)}{\Phi(\phi)}  = 0\ .
\end{equation}
This holds only if
\bea
    \Phi''(\phi) & = & - l^2 \Phi(\phi) \\
    \sin^2 \theta \,\Theta''(\theta) + \tan\theta \,(\sin^2 \theta + 1) \Theta'(\theta) & = & l^2 \Theta(\theta)\ , \label{thetaeq}
\eea
where $l$ is, in general, real or imaginary. However, a negative separation constant would break the $2\pi$ periodicity of $u(\theta,\cdot)$ so we can focus on the case $l\in \mathbb{R}$. The ODEs are then solved by
\bea
    \Phi(\phi) & = & A_1 \sin l\phi + A_2 \cos l\phi \\
    \Theta(\theta) & = & B_1 (1+l c ) \tan^l(\theta/2) + B_2 (1-l c )\tan^{-l}(\theta/2)\ .
\eea
The second linearization of the minimal surface equation with respect to one parameter and a family of solutions thereof are given in Eqs. (3.49) and (3.53) in \cite{Hubeny:2012ry}. Here however we are interested in two-parameter linearization, in which case picking the coefficient of the $\eps_1\eps_2$ term in the minimal surface equation yields
\begin{equation}
    \sin^2\theta  \frac{\pt^2 w}{\pt \theta^2} + \frac{\pt^2 w}{\pt \phi^2} + \tan\theta \,(\sin^2 \theta + 1) \frac{\pt w}{\pt \theta} = \frac{2}{\rho_0}\left( \sin^2\theta \frac{\pt u_1}{\pt\theta}\frac{\pt u_2}{\pt\theta} + \frac{\pt u_1}{\pt\phi}\frac{\pt u_2}{\pt\phi}\right) \ .
\end{equation}
Here $u_1$ and $u_2$ are arbitrary first-order solutions that act as a source for the second-order solution. Notice that this could also be obtained from the one-parameter second linearization by the {polarization formula}. Indeed, if we denote by $W[u]$ as the second order linearization of 
\be
\rho(\theta,\phi) = \rho_0 + \epsilon u(\theta,\phi) + \frac{\epsilon^2}{2} v(\theta,\phi) + {\cal{O}}(\epsilon^3) \ ,
\ee
then the $w(\theta,\phi)$ in \eqref{eq: eps expansion} satisfies
\be
 w(\theta,\phi) = \frac{1}{4}\left( W(u_1+u_2) - W(u_1-u_2)\right)\ .
\ee

\bibliographystyle{JHEP}
\bibliography{ref}

\end{document}